\documentclass[12pt,preprint]{aastex}
\newcommand{\nraoblurb}{The National Radio Astronomy Observatory is
a facility of the National Science Foundation operated under cooperative
agreement by Associated Universities, Inc.}

\newcommand{\arcmper}{\rlap.{^{\prime}}}

\newcommand{\khz}{\ensuremath{\,{\rm kHz}}}
\newcommand{\mhz}{\ensuremath{\,{\rm MHz}}}
\newcommand{\ghz}{\ensuremath{\,{\rm GHz}}}

\newcommand{\K}{\ensuremath{\,{\rm K}}}

\newcommand{\percc}{\ensuremath{\,{\rm cm^{-3}}}}
\newcommand{\kpc}{\ensuremath{\,{\rm kpc}}}

\newcommand{\kms}{\ensuremath{\,{\rm km\, s^{-1}}}}

\newcommand{\s}{\,s}
\newcommand{\microns}{\ensuremath{\, \mu {\rm m}}}
\newcommand{\ev}{\,ev}

\newcommand{\microgauss}{\,\ensuremath{\mu}G}

\newcommand{\cx}[2]{${\rm C}#1#2\alpha$}
\newcommand{\cc}[3]{${\rm C}#1#2#3\alpha$}

\newcommand{\nexpo}[2]{\ensuremath{#1 \times 10^{#2}}}
\newcommand{\hi}{H\,{\sc i}}

\newcommand{\hii}{H\,{\sc ii}}

\newcommand{\ngc}[1]{NGC~#1}

\newcommand\urltilda{\kern -.15em\lower .7ex\hbox{\~{}}\kern .04em}

\newcommand{\lsim}{\ensuremath{\lesssim}}
\begin{document}


\title{Magnetic Field Strengths in Photodissociation Regions}

\author{Dana S. Balser\altaffilmark{1}, D. Anish
  Roshi\altaffilmark{1}, S. Jeyakumar\altaffilmark{2},
  T. M. Bania\altaffilmark{3}, Benjamin T. Montet\altaffilmark{4,5}, \&
  J. A. Shitanishi\altaffilmark{6}}

\altaffiltext{1}{National Radio Astronomy Observatory, 520 Edgemont
  Rd., Charlottesville, VA 22903, USA.}

\altaffiltext{2}{Departamento de Astronom{\'i}a, Universidad de
  Guanajuato, AP 144, Guanajuato CP 36000, Mexico.}

\altaffiltext{3}{Institute  for Astrophysical Research,  Department of
  Astronomy,  Boston University,  725 Commonwealth  Avenue,  Boston MA
  02215, USA.}

\altaffiltext{4}{Harvard-Smithsonian Center for Astrophysics,
  Cambridge, MA 02138 USA.}

\altaffiltext{5}{Cahill Center for Astronomy and Astrophysics, 1200
  E. California Blvd, MC 249-17, Pasadena, CA 91125 USA.}

\altaffiltext{6}{Department of Physics and Astronomy, University of
  Southern California, Los Angeles, CA 90089, USA.}

\begin{abstract}

  We measure carbon radio recombination line (RRL) emission at
  5.3\ghz\ toward four \hii\ regions with the Green Bank Telescope
  (GBT) to determine the magnetic field strength in the
  photodissociation region (PDR) that surrounds the ionized gas.
  Roshi (2007) suggests that the non-thermal line widths of carbon
  RRLs from PDRs are predominantly due to magneto-hydrodynamic (MHD)
  waves, thus allowing the magnetic field strength to be derived.  We
  model the PDR with a simple geometry and perform the non-LTE
  radiative transfer of the carbon RRL emission to solve for the PDR
  physical properties.  Using the PDR mass density from these models
  and the carbon RRL non-thermal line width we estimate total magnetic
  field strengths of $B \sim 100-300$\microgauss\ in W3 and
  \ngc{6334A}.  Our results for W49 and \ngc{6334D} are less well
  constrained with total magnetic field strengths between $B \sim\
  200-1000$\microgauss.  \hi\ and OH Zeeman measurements of the
  line-of-sight magnetic field strength ($B_{\rm los}$), taken from
  the literature, are between a factor of $\sim 0.5-1$ of the lower
  bound of our carbon RRL magnetic field strength estimates.  Since
  $|B_{\rm los}| \le B$, our results are consistent with the magnetic
  origin of the non-thermal component of carbon RRL widths.

\end{abstract}


\section{Introduction}\label{sec:intro}

Magnetic fields play an important role in many astrophysical objects
including planets, stars, and galaxies \citep{parker79}.  Measurements
of magnetic fields in the cosmos, however, are difficult and therefore
their paucity limits our ability to fully understand a wide range of
astrophysical processes.  For example, the role of magnetic fields in
star formation is currently a hotly debated topic \citep[see][and
references within]{crutcher12}.  There are a handful of magnetic field
diagnostics such as dust polarization, Faraday Rotation, and the
Zeeman effect.  Only the Zeeman effect can directly measure the
line-of-sight magnetic field strength in interstellar clouds
\citep{crutcher12}.

Observations of spectral lines from molecular, neutral, and ionized
gas indicate line widths that are broader than the thermal width, even
on small spatial scales where macroscopic effects such as rotation
would be minimal.  These non-thermal line widths are thought to be a
result of motions either from MHD waves \citep[e.g.,][]{mouschovias75}
or turbulence \citep[e.g.,][]{morris74}.  Specific examples include CO
in molecular clouds \citep{arons75} and H$\alpha$ in \hii\ regions
\citep{ferland01, beckman04}.  H$_{2}$CO absorption toward compact
extragalactic sources reveals secular changes to the absorption
intensity on AU scales, with a non-thermal component to the velocity
dispersion, indicating that these motions occur on very small spatial
scales \citep{marscher93}.

Carbon radio recombination line (RRL) emission detected toward star
formation complexes originates from the cooler, mostly neutral gas,
surrounding the \hii\ region called the photodissociation region
(PDR).  \citet{roshi07} suggested that the non-thermal line widths of
carbon RRLs toward PDRs adjacent to \hii\ regions are dominated by
MHD waves and could be used to derive the magnetic field strength.
The PDR is a thin layer lying between the molecular cloud and the
\hii\ region.  At cm-wavelengths the carbon RRL intensity is enhanced
by stimulated emission from the background \hii\ region
\citep{roshi05, quireza06}.  This provides information about the
geometry and allows for relatively simple PDR models to be developed
\citep[see][]{roshi05}.  Here we test the \citet{roshi07} hypothesis
by measuring the non-thermal line widths in four PDRs that also have
either \hi\ or OH Zeeman based determinations of magnetic field
strength.

\section{Observations and Data Reduction}\label{sec:obs}

We observed the RRL and continuum emission at C-band (4-6\ghz) toward
four \hii\ regions with the National Radio Astronomy Observatory
(NRAO)\footnote{\nraoblurb} Green Bank Telescope (GBT) on 5 April, 27
April, and 8 May 2008.  The GBT has a half-power beam-width (HPBW) of
$2\arcmper34$ at an observing frequency of 5.3\ghz.  The aperture
efficiency and beam efficiency are 0.70 and 0.92, respectively,
yielding a sensitivity of 2 K/Jy.  We selected W3, W49, \ngc{6334A},
and \ngc{6334D} as targets since these sources have bright carbon RRL
emission regions and Zeeman measurements in the neutral gas.
Moreover, these sources have been detected in carbon RRLs with a
similar spatial resolution but at a different frequency (8.7 GHz) with
the NRAO 140 Foot telescope \citep{quireza06}.  We require at least
two carbon RRLs, separated in frequency, to model the PDR and derive
the magnetic field strength (see \S{\ref{sec:model}}).  The \cx91\ and
\cx92\ RRLs were observed with the 140 Foot.  At these frequencies the
140 Foot has a HPBW of $3\arcmper20$.  The aperture efficiency and
beam efficiency are 0.51 and 0.68, respectively, giving 0.27 K/Jy.

For each GBT observing session we first checked the pointing and focus
by observing a nearby calibrator.  Continuum scans in R.A. and
Decl. were then made to measure the free-free emission.  We scanned
the GBT both forward and backward for each cardinal direction while
simultaneously sampling both linear, orthogonal polarizations.  So
each continuum observation consisted of 4 scans times 2 polarizations
or 8 total measurements.  We used the Digital Continuum Receiver (DCR)
with a bandwidth of 80\mhz, centered at 5.3\ghz, and an integration
time of 0.1\s.  The GBT was driven at a rate of 80 arcsec per second
for 30\s, providing a scan length of 40 arcmin.  Data from both
directions (forward and backward) and linear polarizations (XX and YY)
were averaged for several continuum observations to increase the
signal-to-noise ratio of the continuum intensity measurement.

Finally, spectra were taken using total power, position switching
where we observed a reference (OFF) position for 6 minutes, and then,
tracking the same sky path, observed the target (ON) position for 6
minutes.  The Autocorrelator Spectrometer (ACS) was configured with 8
spectral windows, each with two orthogonal, linear polarizations
yielding 16 independent spectra.  Each spectral window contained 4096
channels with a bandwidth of 12.5\mhz, providing a spectral resolution
of 3.05\khz, or 0.18\kms\ per channel at 5\ghz.  We centered each
spectral window to include the carbon RRLs: \cc104-\cc110, and \cc112.
The \cc111\ transition is confused by a higher order RRL and therefore
was not observed.  The intensity scale was calibrated in Kelvins using
noise diodes that injected noise into the signal path.  We verified
that the accuracy of the calibration was within 10\% by making 
continuum observations toward 3C286.

Both the spectral line and continuum data were reduced and analyzed
with the single-dish software package TMBIDL\footnote{See
  https://github.com/tvwenger/tmbidl.git.}.  Typically, a third-order
polynomial function was fit to the continuum baseline and removed from
the data.  A Gaussian profile was fit to the main continuum source to
determine the peak intensity, full-width at half-maximum (FWHM) \hii\
region size, and the center position.  Spectral line data were reduced
by first averaging spectra in each spectral window, and then combining
the different Cn$\alpha$ transitions to improve the signal-to-noise
ratio.  At these high principal quantum numbers the difference in
energy between adjacent Cn$\alpha$ transitions is negligible and
therefore we can average these different transitions
\citep[e.g.,][]{balser06}.  This was done by first resampling the
spectral channels of each spectral window to match the velocity
resolution of the 104$\alpha$ spectral window and then shifting each
spectrum to be at the same LSR velocity.  Here no correction was made
for the different HPBW's.  The \cc112\ RRL was not included in the
average, however, because of variations in the calibration scale near
the carbon RRL that we suspect was caused by resonances in the
telescope feed.  A third-order polynomial was fit to the line-free
regions of the spectral baseline to remove the continuum level and any
other instrumental baseline structure.  Multiple Gaussian functions
were then fit to the various RRLs within each spectral window to
determine the peak line intensity, the FWHM line width, and the LSR
velocity.  The He and heavier element RRLs were fit simultaneously,
whereas the H RRL was fit separately.

\section{Radio Continuum and RRL Results}\label{sec:results}

\begin{figure}
\includegraphics[angle=270,scale=0.32]{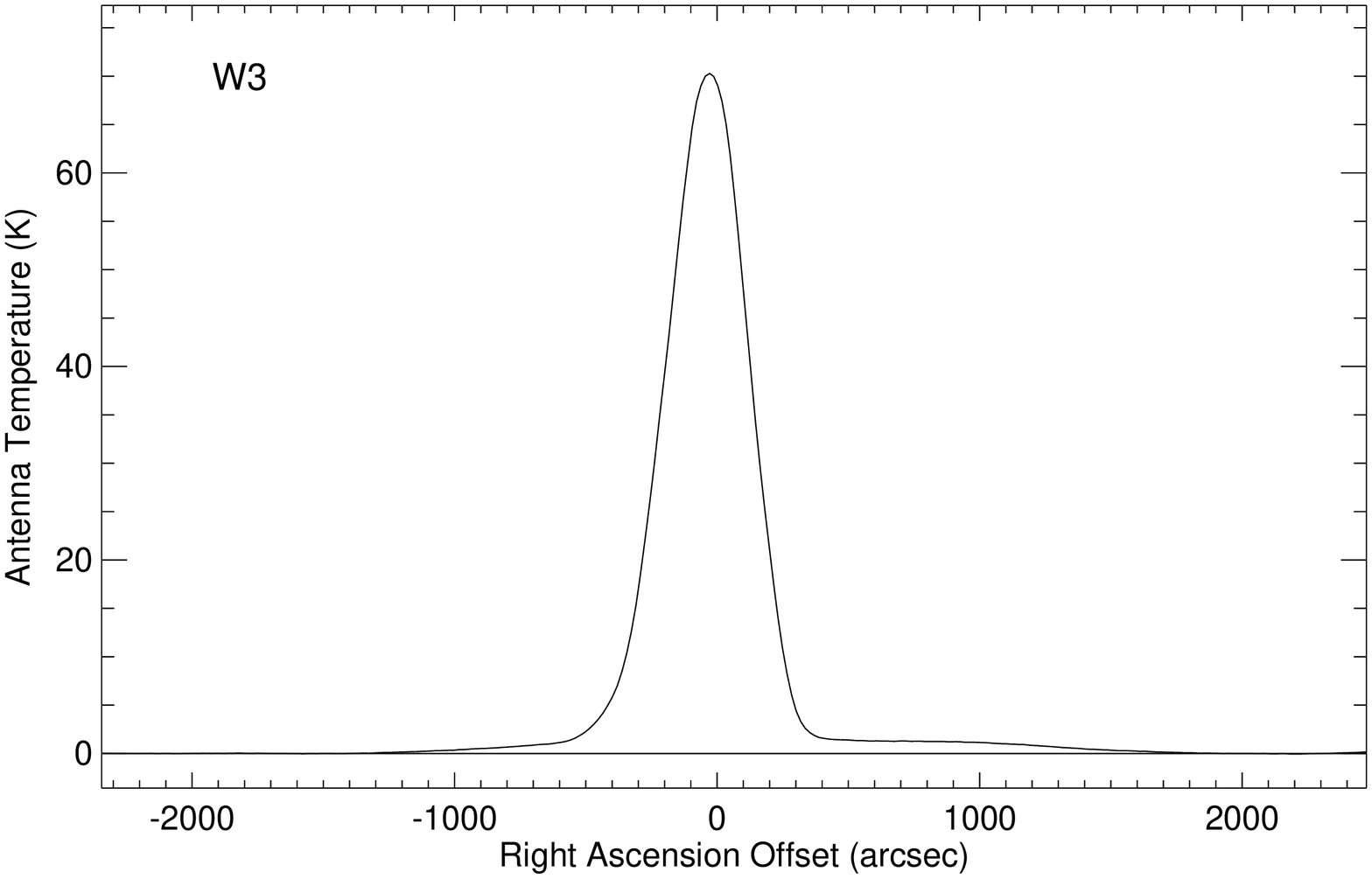} 
\includegraphics[angle=270,scale=0.32]{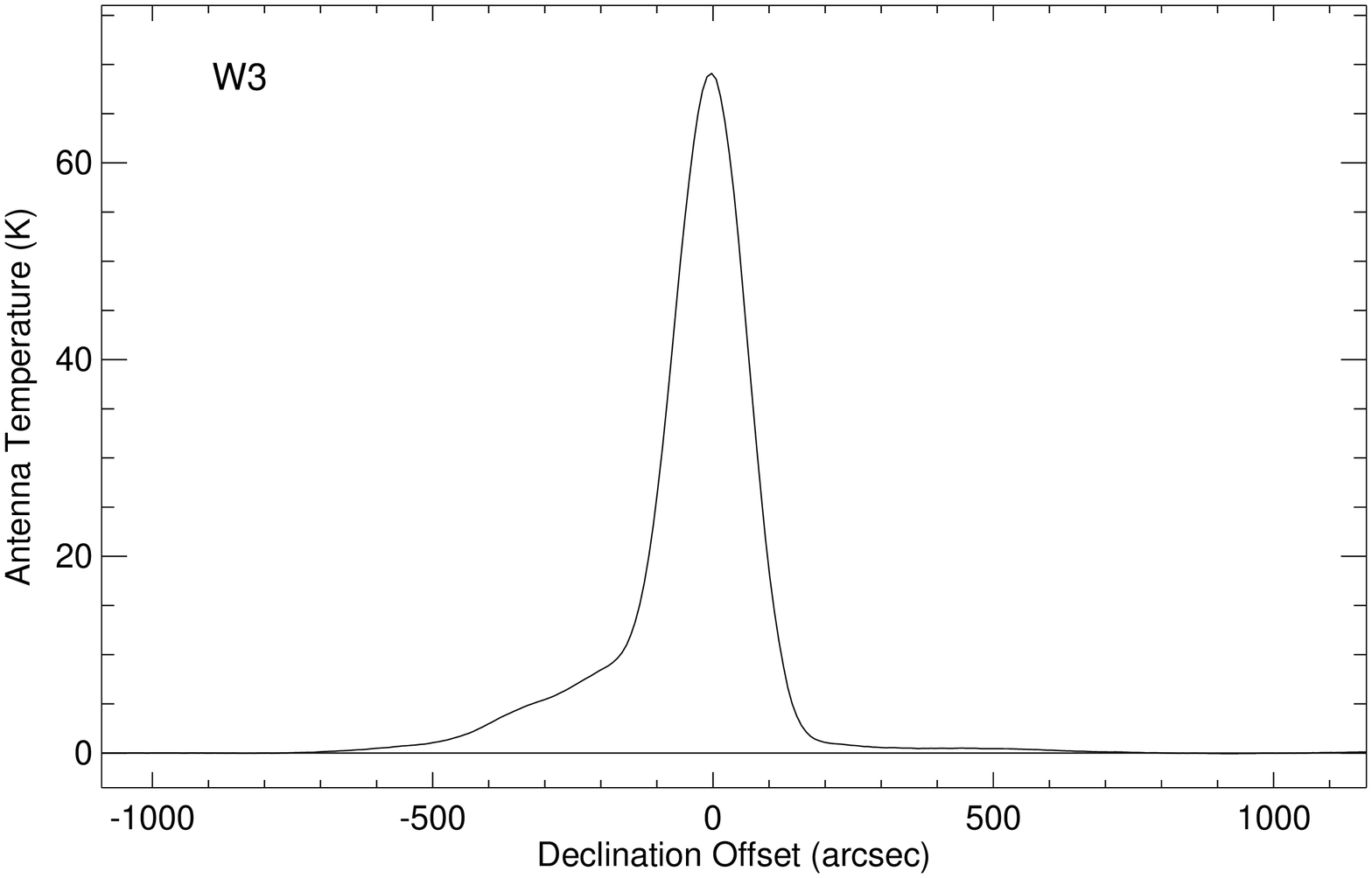} 
\includegraphics[angle=270,scale=0.32]{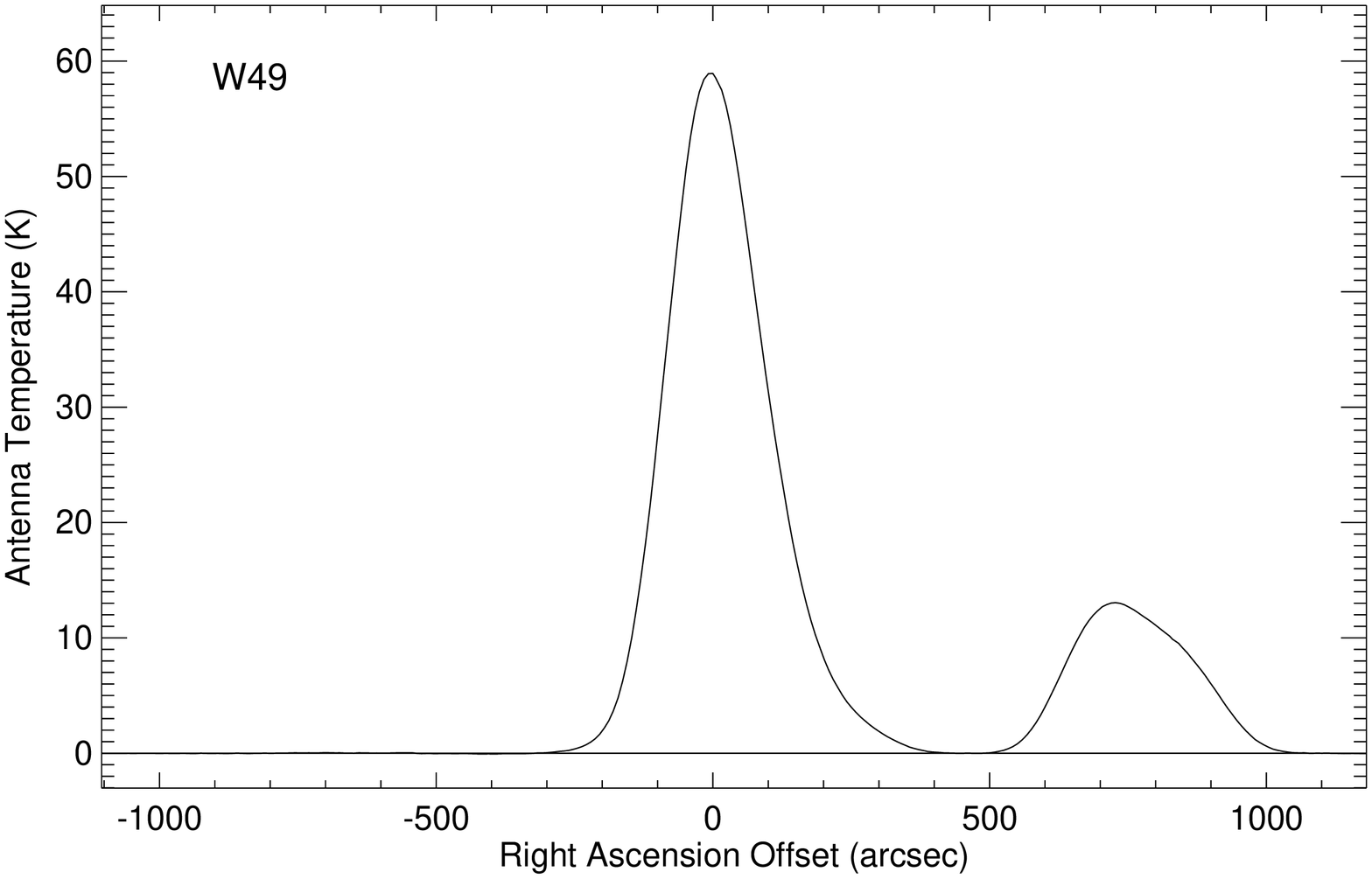} 
\includegraphics[angle=270,scale=0.32]{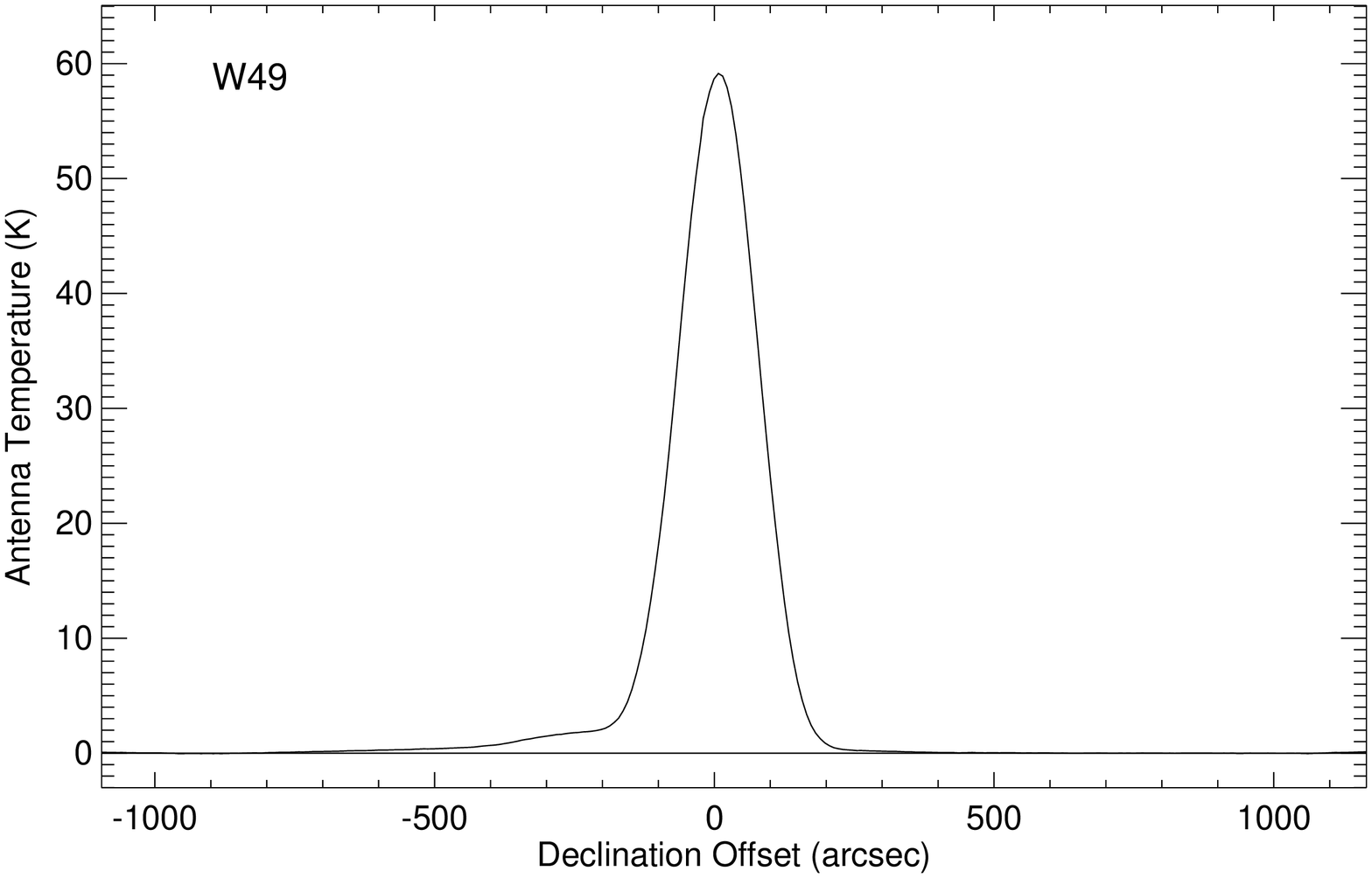} 
\caption{Continuum data for W3 (top) and W49 (bottom).  The antenna
  temperature is plotted as a function of offset position relative to
  the nominal coordinates in Table~\ref{tab:cont} for the R.A. scan
  (left) and the Decl. scan (right).  A polynomial has been fit to the
  baseline to remove any instrumental effects such as weather.}
\label{fig:cont_w}
\end{figure}

\begin{figure}
\includegraphics[angle=270,scale=0.32]{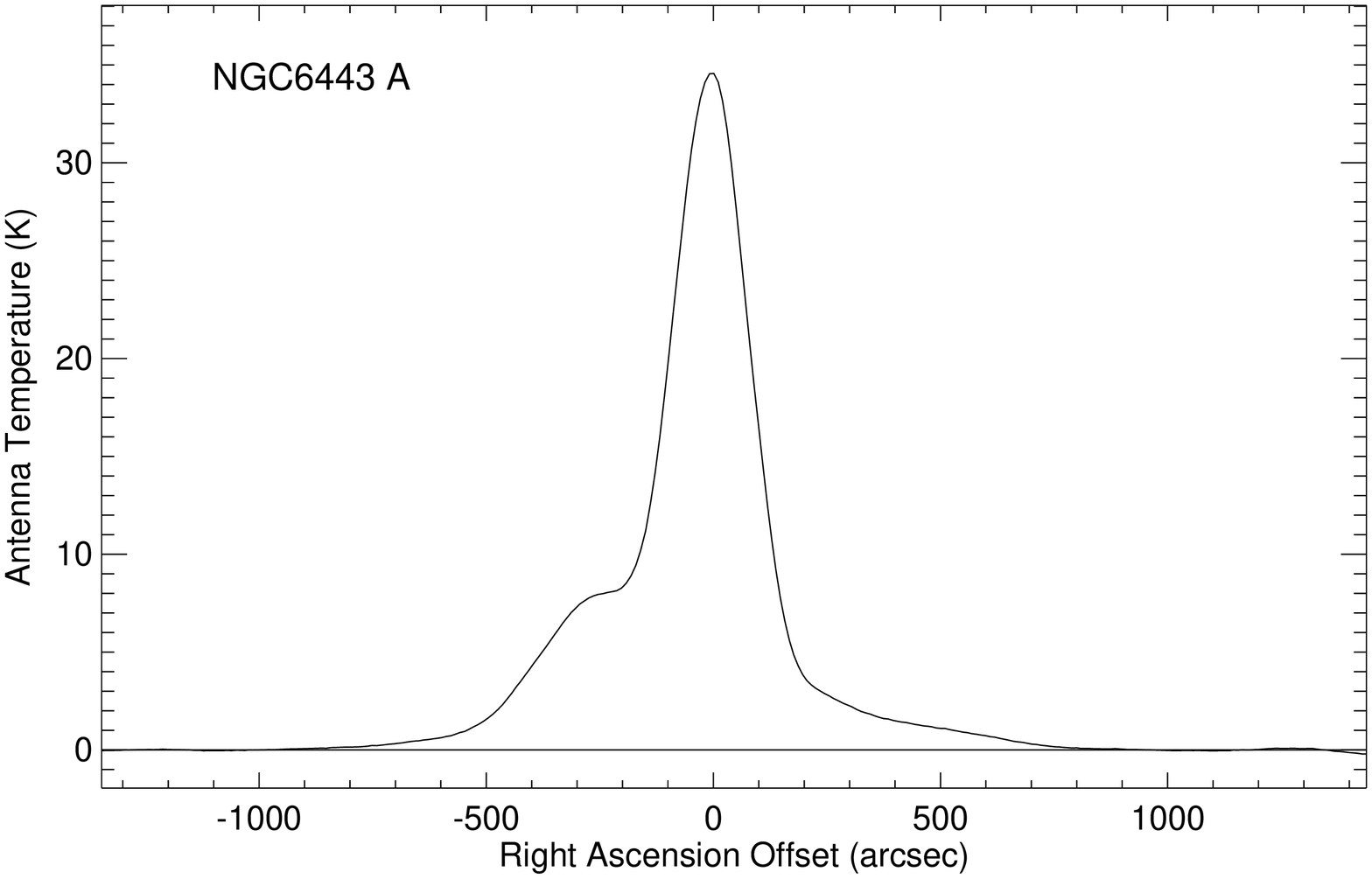} 
\includegraphics[angle=270,scale=0.32]{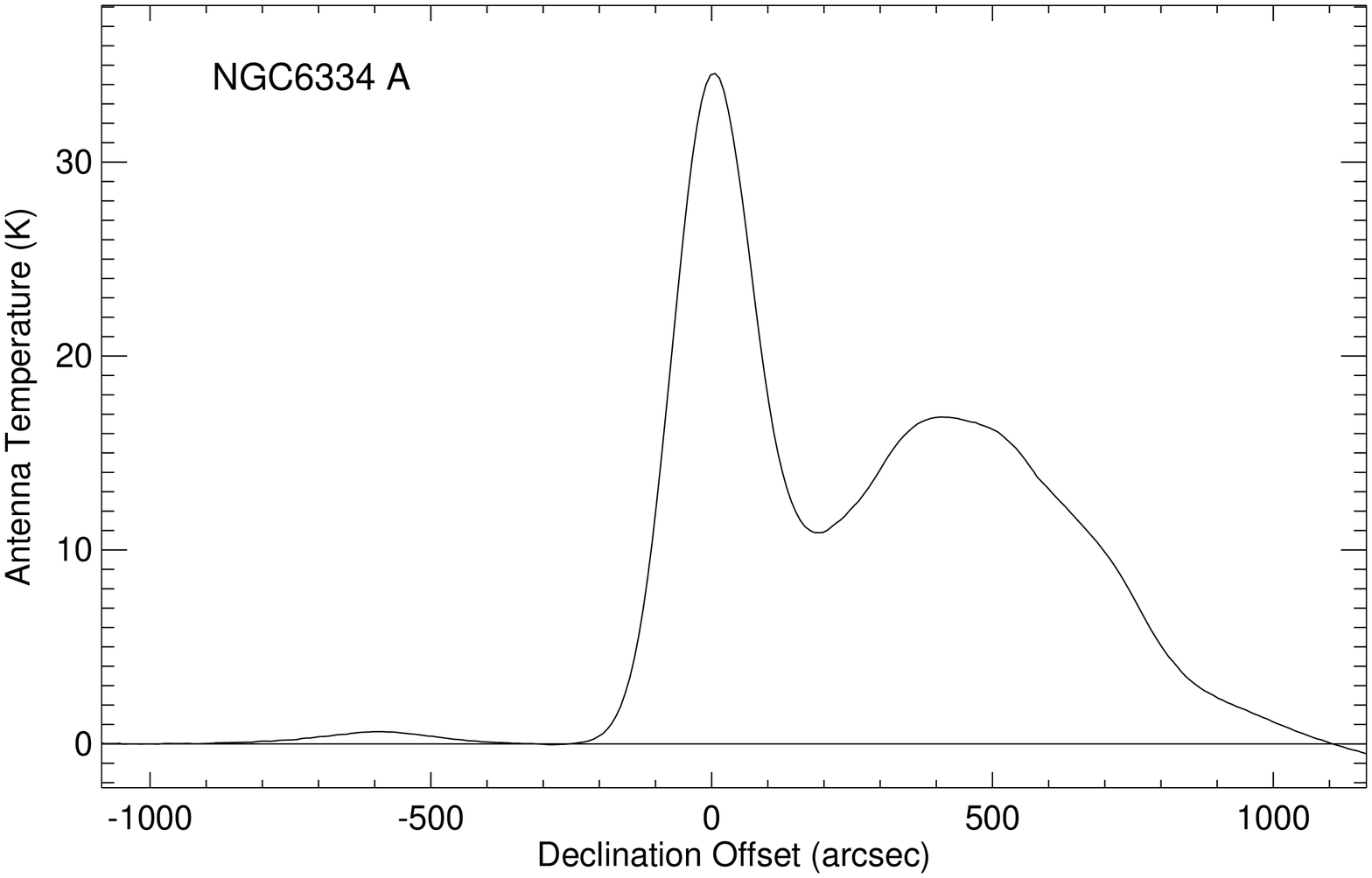} 
\includegraphics[angle=270,scale=0.32]{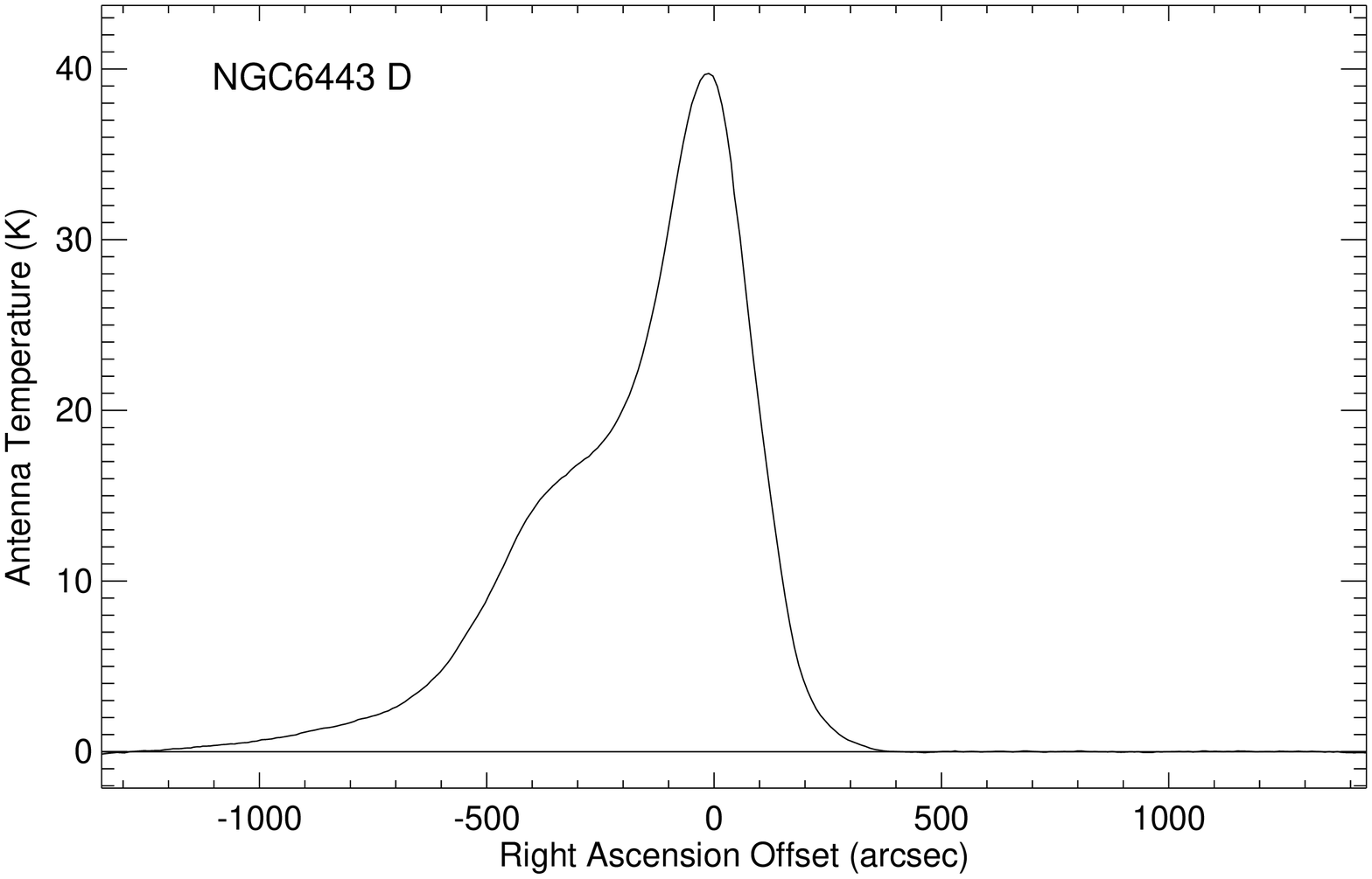} 
\includegraphics[angle=270,scale=0.32]{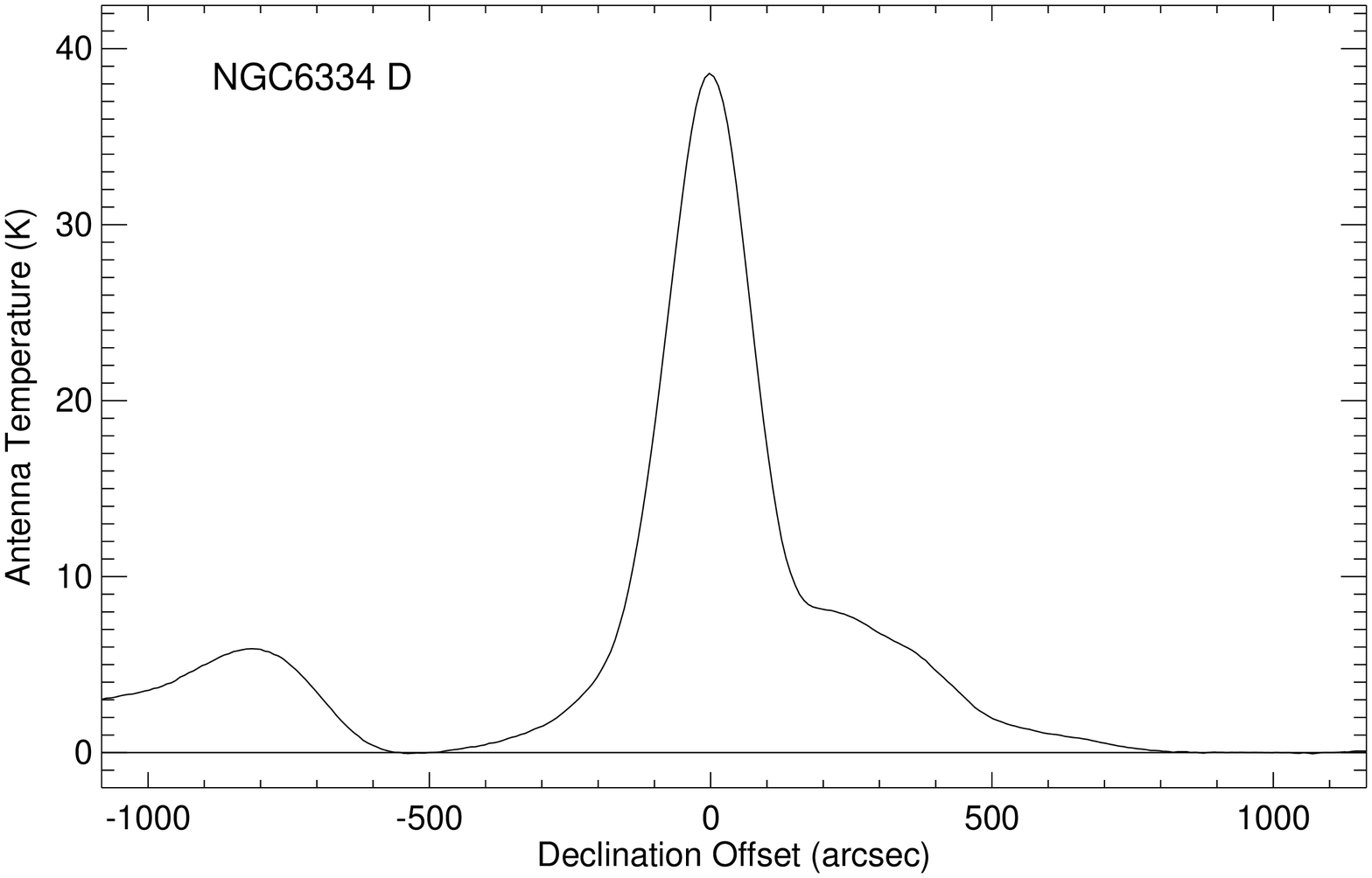} 
\caption{Continuum data for \ngc{6334A} (top) and \ngc{6334D}
  (bottom).  See Figure~\ref{fig:cont_w} for details.}
\label{fig:cont_ngc}
\end{figure}

Star forming complexes that contain early-type stars consist of \hii\
regions that have formed due to the large number of hydrogen-ionizing
photons, molecular clouds where the next generation of stars may form,
and PDRs that lie at their interface.  The radio continuum emission
observed toward star forming complexes is primarily produced from
free-free emission in the \hii\ region.  The non-thermal Galactic
background emission may contribute to the observed continuum, but
because this background emission is smoothly distributed over spatial
scales larger than the \hii\ region size it will be removed in our
baseline fitting procedures.
Figures~\ref{fig:cont_w}-\ref{fig:cont_ngc} show the continuum
profiles for both the R.A. and Decl. scans.  Table~\ref{tab:cont}
summarizes the \hii\ region continuum parameters based on Gaussian
fits to the main source component.  Listed are the source name, the
B1950 equatorial coordinates, the distance from the Sun, $R_{\rm
  Sun}$, and the peak intensity, $T_{\rm C}$, and FWHM size, $\Theta$,
and their associated $1\sigma$ errors.

\begin{figure}
\includegraphics[angle=270,scale=0.32]{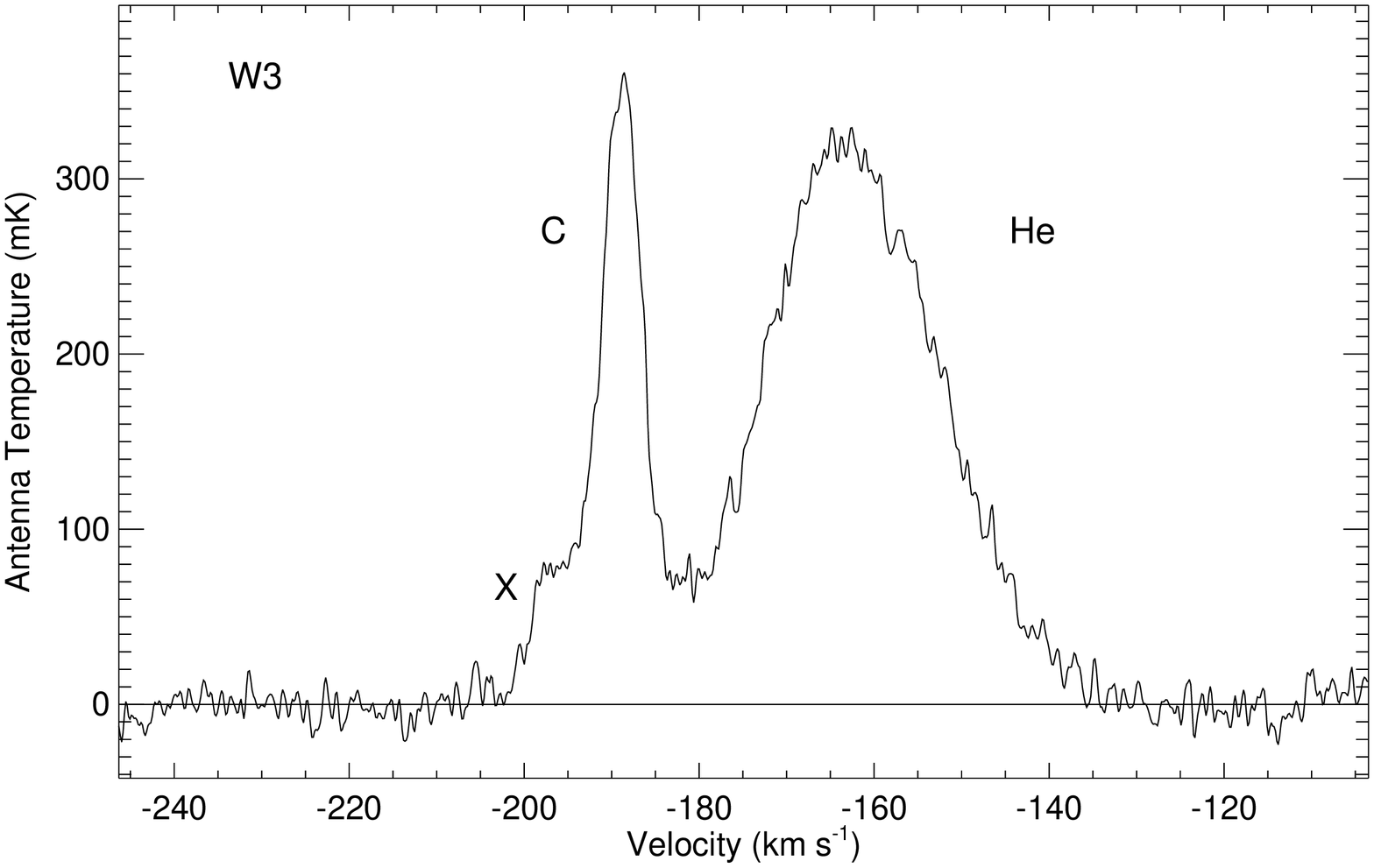} 
\includegraphics[angle=270,scale=0.32]{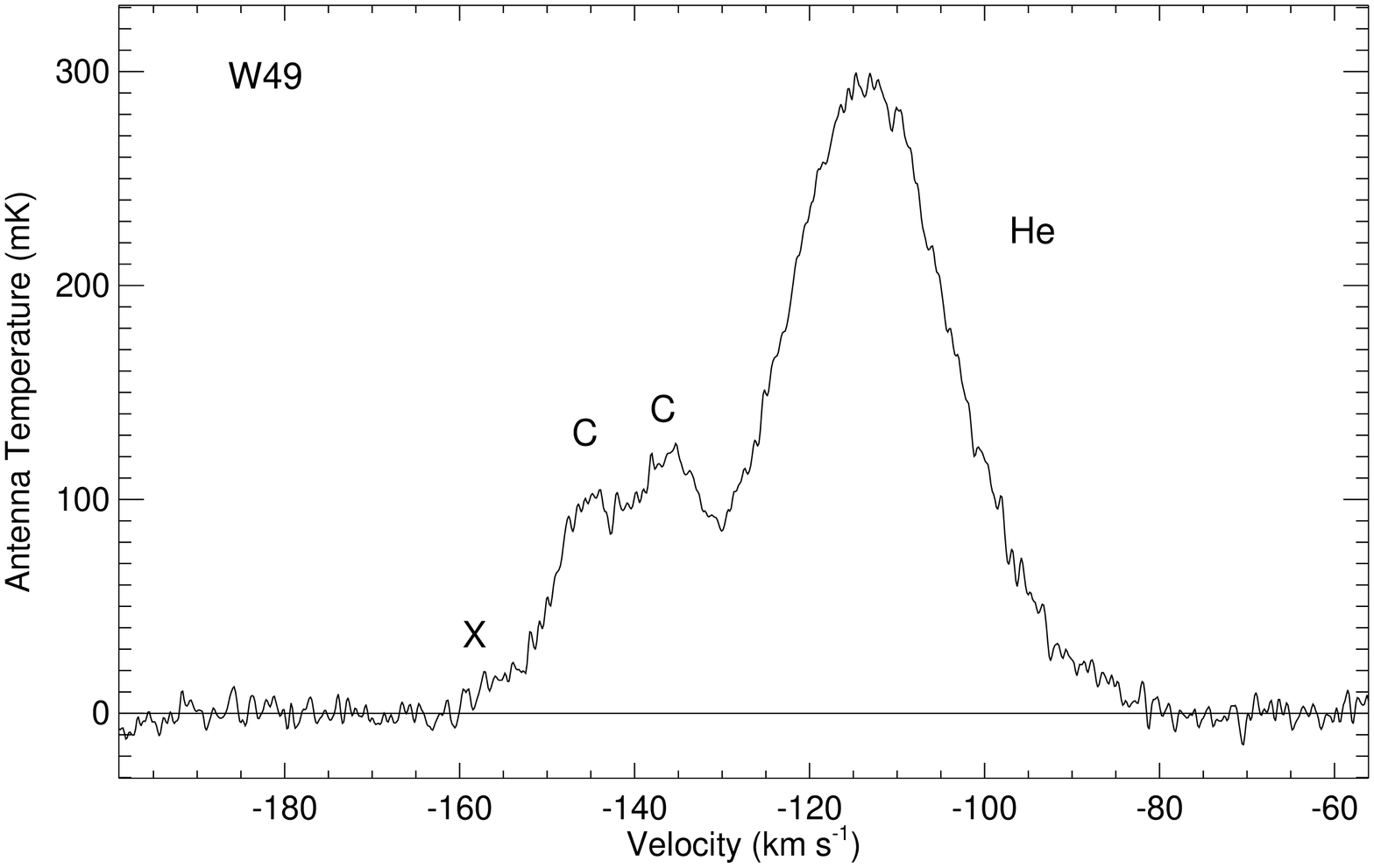} 
\includegraphics[angle=270,scale=0.32]{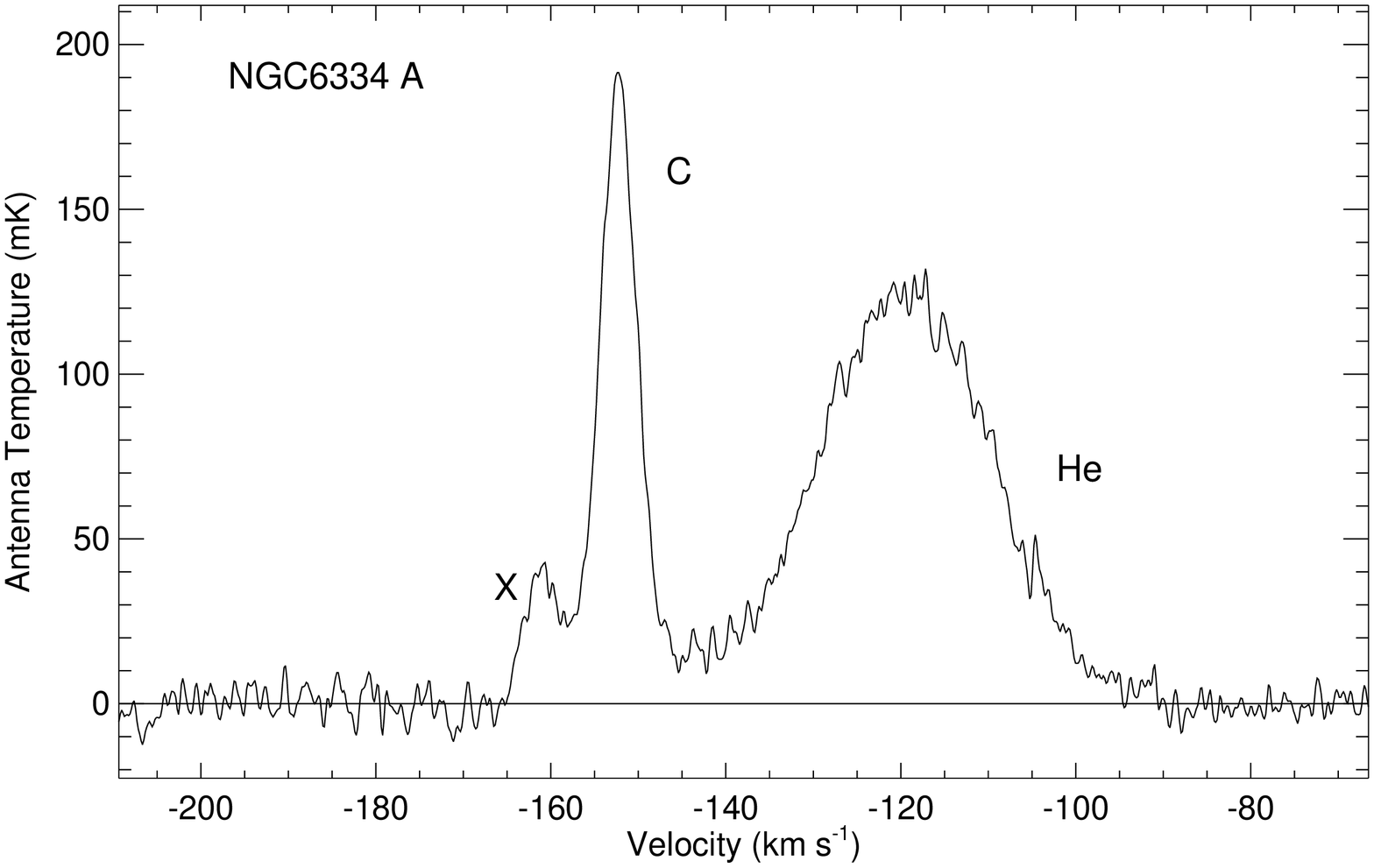} 
\includegraphics[angle=270,scale=0.32]{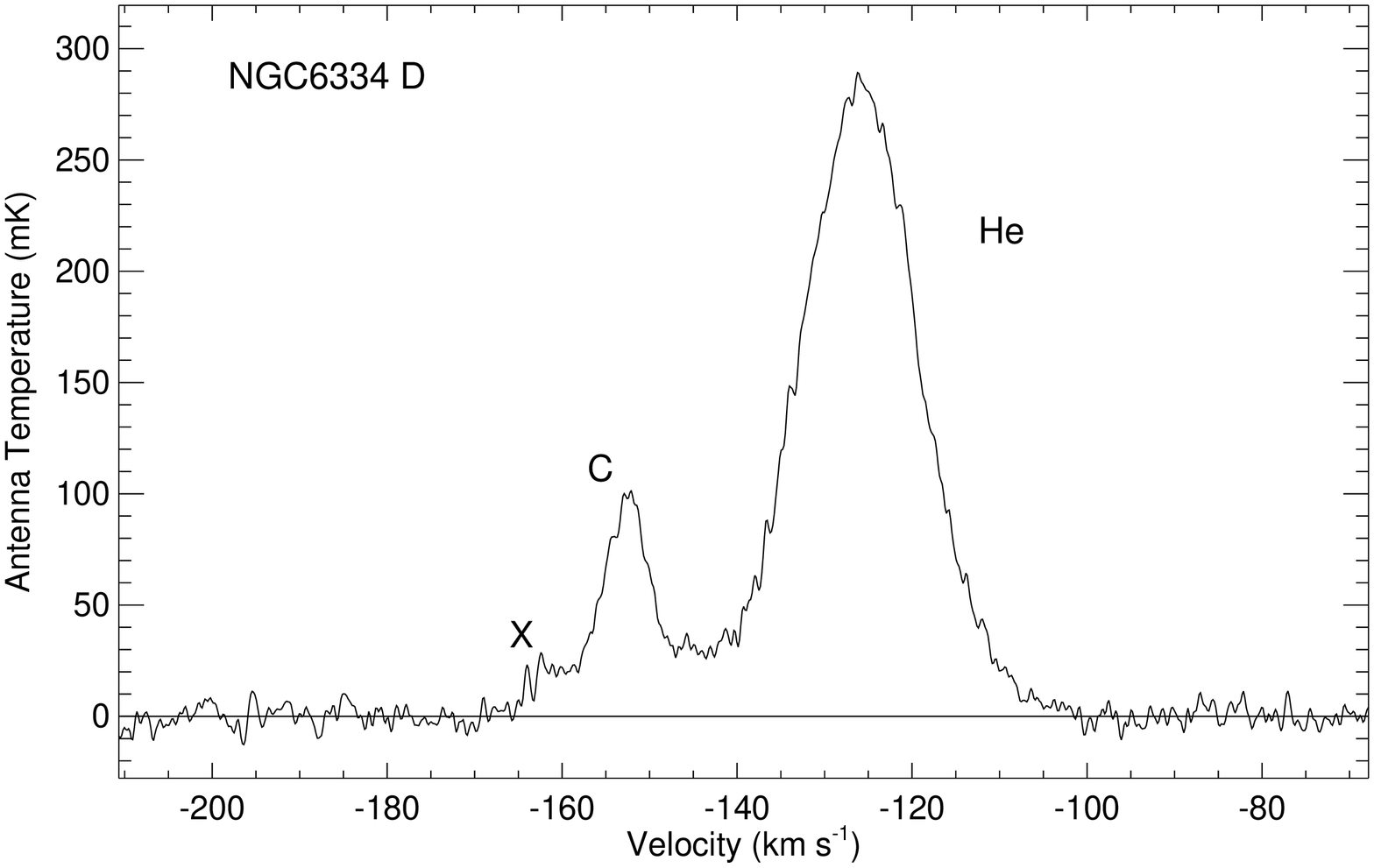} 
\caption{\hii\ regions radio recombination line spectra at 5.3\ghz.
  The antenna temperature is plotted as a function of LSR velocity
  relative to the hydrogen RRL.  The view has been magnified to show
  only the helium and heavier element RRLs.  The continuum emission
  and any instrumental baseline has been removed using a third-order
  polynomial fit to the line-free regions.  The horizontal solid line
  defines the zero-level.  The helium and carbon lines are labeled.
  The label ``X'' denotes that the line identification is uncertain
  (see text).}
\label{fig:line}
\end{figure}

We detect hydrogen and helium RRL emission from the four \hii\ regions
in our sample.  Typically RRLs from heavier elements are not detected
from \hii\ regions since they have small abundances producing line
intensities below the sensitivity limit of most radio telescopes.  In
many cases, however, a narrower, weaker line is detected at higher
frequencies and has been identified as carbon RRL emission formed
within the PDR \citep[e.g., see][]{zuckerman68, wenger13}.  The
physical temperature of PDRs is about an order of magnitude lower than
that in \hii\ regions.  This lower temperature makes the carbon RRL
from PDRs detectable since the line optical depth has a strong inverse
dependence on the gas temperature.  Since the carbon RRL arises from
the PDR it often has a slightly different LSR velocity than the
hydrogen and helium RRLs.  Figure ~\ref{fig:line} shows spectra for
each source with a magnified view to highlight the carbon profiles.
The velocity scale is defined relative to the hydrogen RRL which
resides about 150\kms\ at more positive velocities.  Each spectrum
reveals multiple heavy element RRL profiles.  Carbon is likely to be
the brightest heavy element RRL because of its low ionization
potential (11.3\ev), high cosmic abundance, and low depletion.  Other
candidates are sulfur and magnesium.  The W49 spectrum contains two
carbon RRLs that have been shown to originate from spatially distinct
PDRs \citep{roshi06}.  For each source there exists a weaker
transition, labeled as ``X'', that is consistent with an element
heavier than carbon since the line center is at higher frequencies.
For W3 and W49 the ``X'' line may be another carbon RRL from a
different component, or possibly sulfur.  Based on the center velocity
and reduced line intensity we expect the ``X'' line to be sulfur in
the two \ngc{6334} sources.  Table~\ref{tab:line} summarizes the RRL
line parameters. Listed are the source name, the element, the peak
intensity, $T_{\rm L}$, the FWHM line width, $\Delta{V}$, the LSR
velocity, $V_{\rm LSR}$, the total integration time, $t_{\rm integ}$,
and root-mean-square noise in the line-free region, rms, together with
their $1\,\sigma$ errors.

\section{Photodissociation Region Models}\label{sec:model}

To derive the total magnetic field strength requires knowing both the
carbon RRL non-thermal velocity width and the PDR density (see
\S{\ref{sec:b}}). Here we develop PDR models to determine the PDR
density that are constrained by carbon RRLs at two different
frequencies \citep[see, e.g., ][]{roshi05}.  Infrared observations of
\hii\ region/PDR/molecular cloud complexes typically find $22\,\mu m$
emission surrounded by $12\,\mu m$ emission \citep[e.g.,
see][]{anderson14}.  The $22\,\mu m$ emission is produced by
stochastically heated small dust grains within the \hii\ region,
whereas the $12\,\mu m$ emission is thought to be polycyclic aromatic
hydrocarbon, PAH, emission within the PDR.  We therefore consider a
simplified model consisting of homogeneous cylinders of PDR material,
co-located with the molecular gas, in front of the \hii\ region.  This
geometry is consistent with both observations and models.
Observations show that the carbon RRL intensity is correlated with the
\hii\ region continuum intensity \citep[e.g.,][]{quireza06}.  This
implies that the carbon RRL is amplified by stimulated emission from
the \hii\ region which lies behind the PDR.  Models of the radiative 
transfer show that without stimulated emission, at cm-wavelengths, we
would not have the sensitivity to detect the carbon RRL emission
\citep{roshi05}.

We use the formulation of \citet{shaver75} to perform the radiative
transfer and to calculate the carbon RRL brightness temperature at
frequency $\nu$:
\begin{eqnarray}\label{eq:tl}
T_{\rm L}^{\rm B}(\nu) & = & T_{\rm bg}(\nu) e^{-\tau_{C}(\nu)}(e^{-\tau_{\rm L}(\nu)} - 1) + \nonumber \\
                       &   & T_{\rm PDR}\left( \frac{b_{\rm m} \tau_{\rm L}(\nu)^{*} + \tau_{\rm C}(\nu)}{\tau_{\rm L}(\nu) + \tau_{\rm C}(\nu)} (1 - e^{-(\tau_{\rm L}(\nu) + \tau_{\rm C}(\nu))}) - (1 - e^{-\tau_{\rm C}(\nu)})\right),
\end{eqnarray}
where the first term is the contribution to the line temperature due
to the background radiation field and the second term is the intrinsic
emission from the PDR cylinder.  The background temperature, $T_{\rm
  bg}(\nu)$, is dominated by the \hii\ region and so
\begin{equation}\label{eq:bg}
T_{\rm bg}(\nu) = T_{\rm e}(1 - e^{-\tau_{\rm C}^{HII}(\nu)}),
\end{equation}
where $\tau_{C}^{HII}(\nu)$ is the \hii\ region continuum optical
depth given by Equation (31) in \citet{shaver75}.  The PDR thermal
temperature, $T_{\rm PDR}$, is typically between $100-500$\K\
\citep[e.g.,][]{abel05}.  The line and continuum optical depths of the
PDR are given by $\tau_{\rm L}(\nu)$ and $\tau_{\rm C}(\nu)$,
respectively.  We calculate the PDR continuum opacity from Equation
(31) in \citet{shaver75}.

The PDR line opacity is:
\begin{equation}\label{eq:tau}
\tau_{\rm L}(\nu)  =  b_{\rm n} \beta_{\rm n} \tau_{\rm L}(\nu)^{*},
\end{equation}
where $b_{\rm n}$ and $\beta_{\rm n}$ are the departure coefficients
of the energy level $\rm {n}$.  The LTE line opacity $\tau_{\rm
  L}(\nu)^{*} \propto (n_{\rm e}^{\rm PDR} \, n_{\rm i}^{\rm PDR} \,
\ell$) where $n_{\rm e}^{\rm PDR}$ and $n_{\rm i}^{\rm PDR}$ are the
electron and ion number densities of the PDR, respectively, and $\ell$
is the PDR cylinder thickness \citep[see Equation 71 in][]{shaver75}.  We
assume all of the ions in the PDR arise from carbon and therefore
$n_{\rm i}^{\rm PDR} = n_{\rm C^{+}}^{\rm PDR} = n_{\rm e}^{\rm PDR}$.
The departure coefficients are calculated using a new computer code
developed by \citet{roshi14} that includes modification of the level
population of the carbon atom due to a dielectronic-like recombination
process \citep{walmsley82} and a background radiation field from an
\hii\ region.  This is a modified version of the original code
developed by \citet{brocklehurst77} and \citet{walmsley82}.  For the
computation of $b_{\rm n}$ and $\beta_{\rm n}$, we assume 25\% of the
carbon atoms are depleted onto dust grains \citep{natta94}, and a
cosmic carbon abundance of C/H = \nexpo{3.9}{-4} by number
\citep{morton74}.  With these assumptions the hydrogen number density
in the PDR is $n_{\rm H} = 3.4 \times 10^{3} \, n_{\rm e}$.

The PDR models require the background, \hii\ region, intensity as a
function of frequency to be known, for calculating the departure
coefficients.  We adopt the spherical, homogeneous \hii\ region models
of \citet{balser95} and constrain these models with our C-band
(5.3\ghz) radio continuum data, listed in Table~\ref{tab:cont}, to
derive the size, $\theta_{\rm sph}$, electron number density, $n_{\rm
  e}$, and the number of hydrogen-ionizing photons emitted per second,
$N_{\rm L}$. The values of $N_{\rm L}$ provide an estimate of the
stellar spectral type, assuming all of the hydrogen-ionizing photons
come from a single star.  The peak emission measure, $EM = \int{n_{\rm
    e} d\ell}$, is taken from the formalism of \citet{wood89}.  Radio
continuum data alone cannot constrain the electron temperature
($T_{\rm e}$), and therefore we adopt the values from \citet{balser99}
that were derived from RRL and continuum emission at 8.7\ghz.
Table~\ref{tab:hii} lists these physical properties for each \hii\
region in our sample.

We use the numerical code developed by \citet{roshi14} to compute the
carbon RRL flux density from the PDR by solving the non-LTE radiative
transfer equation.  The line temperature, $T_{\rm L}$, provided by the
model is converted to flux density, $S_{\rm L}$, using the equation
\begin{equation}\label{eq:model_flux}
S_{\rm L} = \frac{2\,k\,T_{\rm L}}{\lambda^2}\,\Omega
\end{equation}
where $k$ is the Boltzmann constant, $\lambda$ is the observed
wavelength, and $\Omega$ is the source solid angle.  We assume the
source size equals the GBT's HPBW of $2\arcmper34$.  There are three
free parameters in this model: the PDR temperature, $T_{\rm PDR}$, the
PDR electron number density, $n_{\rm e}^{\rm PDR}$, and the PDR
cylinder thickness, $\ell$.  Since PDR temperatures range from
100-500\K, we consider values of 100\K, 200\K, and 500\K\ for our
models.  The departure coefficients are a function of $T_{\rm PDR}$
and $n_{\rm e}^{\rm PDR}$.  They are calculated for a set of electron
densities between 1 and 500\percc\ for each $T_{\rm PDR}$ value.
Modeling requires two observed carbon RRL intensities to solve for
$n_{\rm e}^{\rm PDR}$ and $\ell$, since we assume a set of values for
the PDR gas temperatures.  Therefore, we use the 140 Foot 8.7\ghz\
carbon RRL data from \citet{quireza06} together with the GBT 5.3\ghz\
observations discussed in \S{\ref{sec:obs}}.  Antenna temperatures are
converted to flux density by using 2 K/Jy and 0.27 K/Jy for the GBT
and 140 Foot, respectively.  We assume the PDR is filling the
$2\arcmper34$ HPBW of the GBT and we therefore scale the 140 Foot flux
density by the ratio of the beam areas,
$(2\arcmper34/3\arcmper20)^{2}$.  Table~\ref{tab:model} summarizes
the constraints to the models.  Listed are the source name, the RRL
transitions, the peak intensity, the FWHM line width, the LSR velocity,
and the flux density, together with their associated errors.  The
errors listed for the carbon RRL peak intensity, FWHM line width, and
LSR velocity are the $1\sigma$ uncertainties in the Gaussian fits to
the line profile.  These errors are propagated to the flux density,
$S_{\nu}$.

\begin{figure}
\includegraphics[angle=0,scale=0.4]{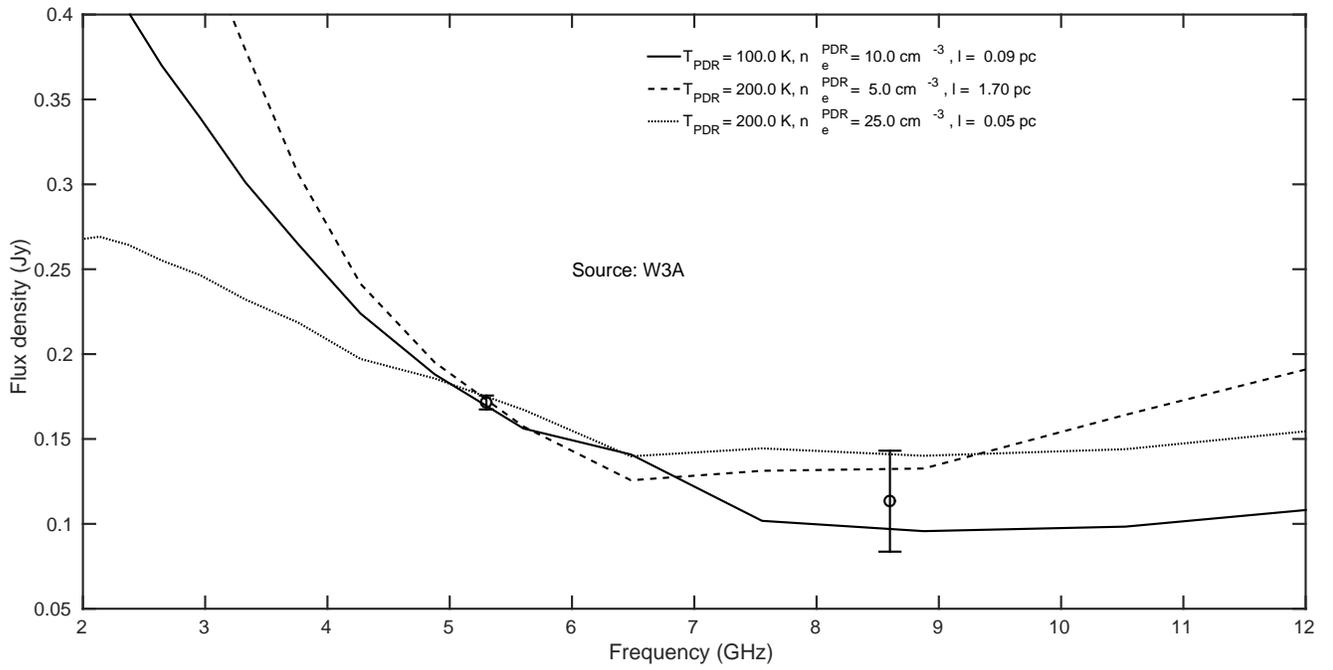} 
\caption{PDR model results for W3.  The flux density is plotted as a
  function of frequency.  The lines correspond to different models and
  the points are from carbon RRL observations at 5.3\ghz\
  (\S{\ref{sec:obs}}) and 8.7\ghz\ \citep{quireza06}.  The error bars
  shown are $\pm\ 3 \sigma$ values.}
\label{fig:model_w3}
\end{figure}

\begin{figure}
\includegraphics[angle=0,scale=0.4]{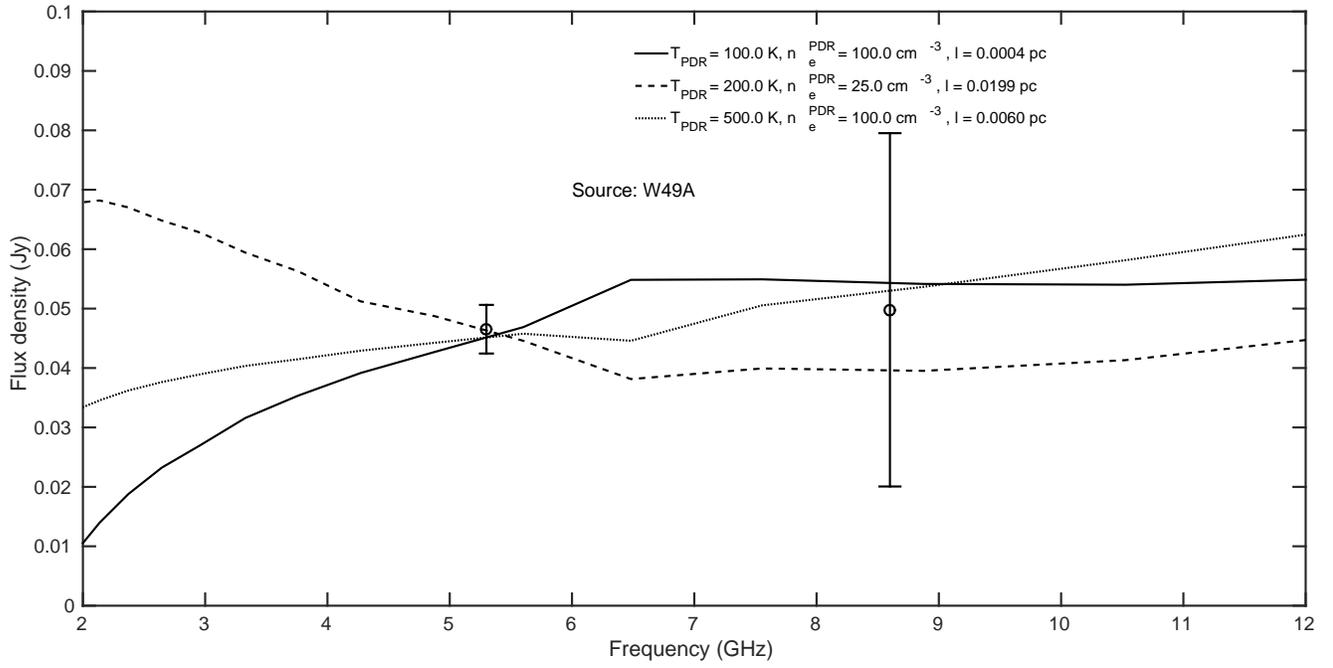} 
\caption{PDR model results for W49.  See Figure~\ref{fig:model_w3} for details.}
\label{fig:model_w49}
\end{figure}

\begin{figure}
\includegraphics[angle=0,scale=0.4]{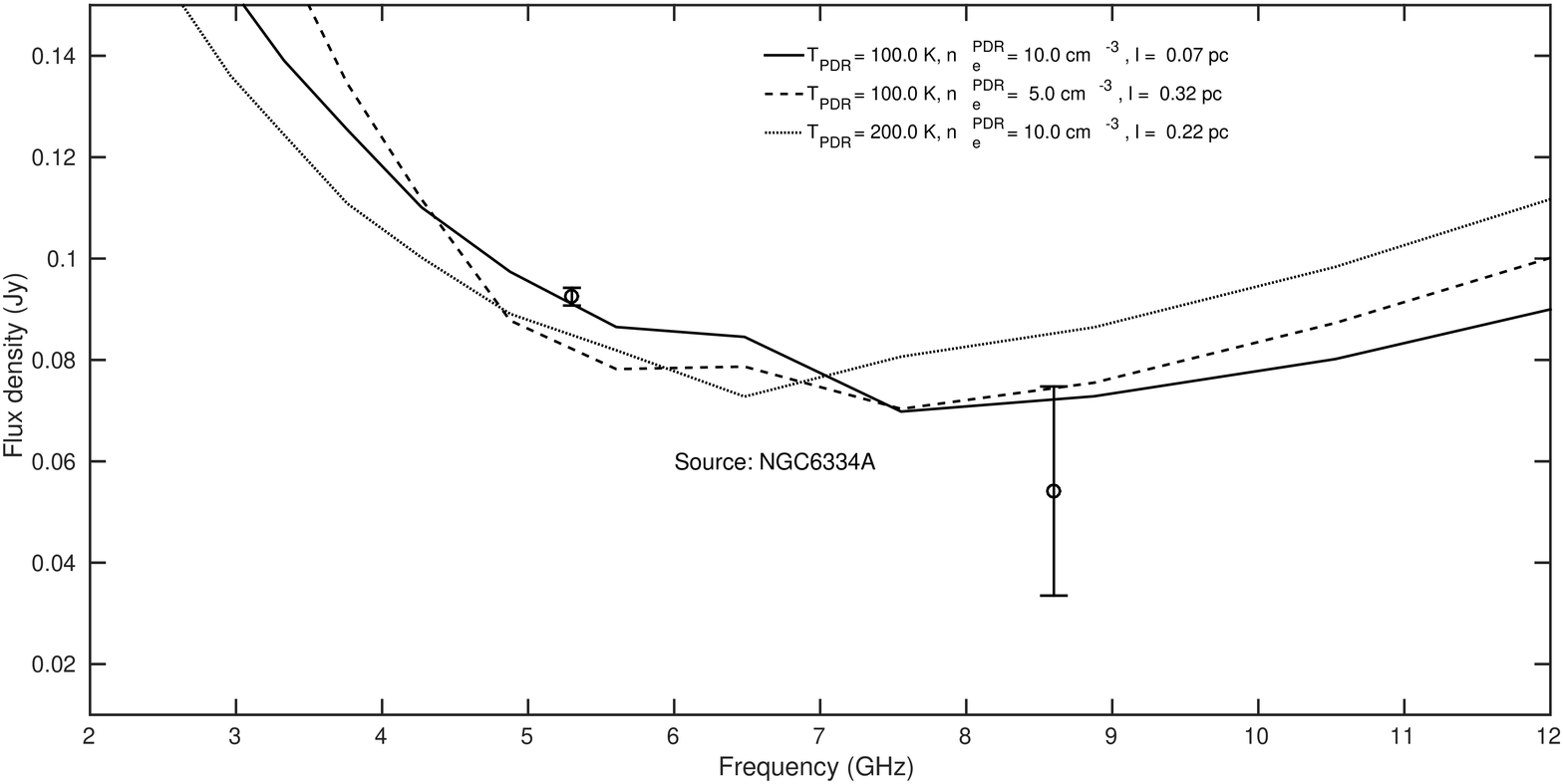} 
\caption{PDR model results for \ngc{6334A}. See Figure~\ref{fig:model_w3} for
  details.}
\label{fig:model_n6334a}
\end{figure}

\begin{figure}
\includegraphics[angle=0,scale=0.4]{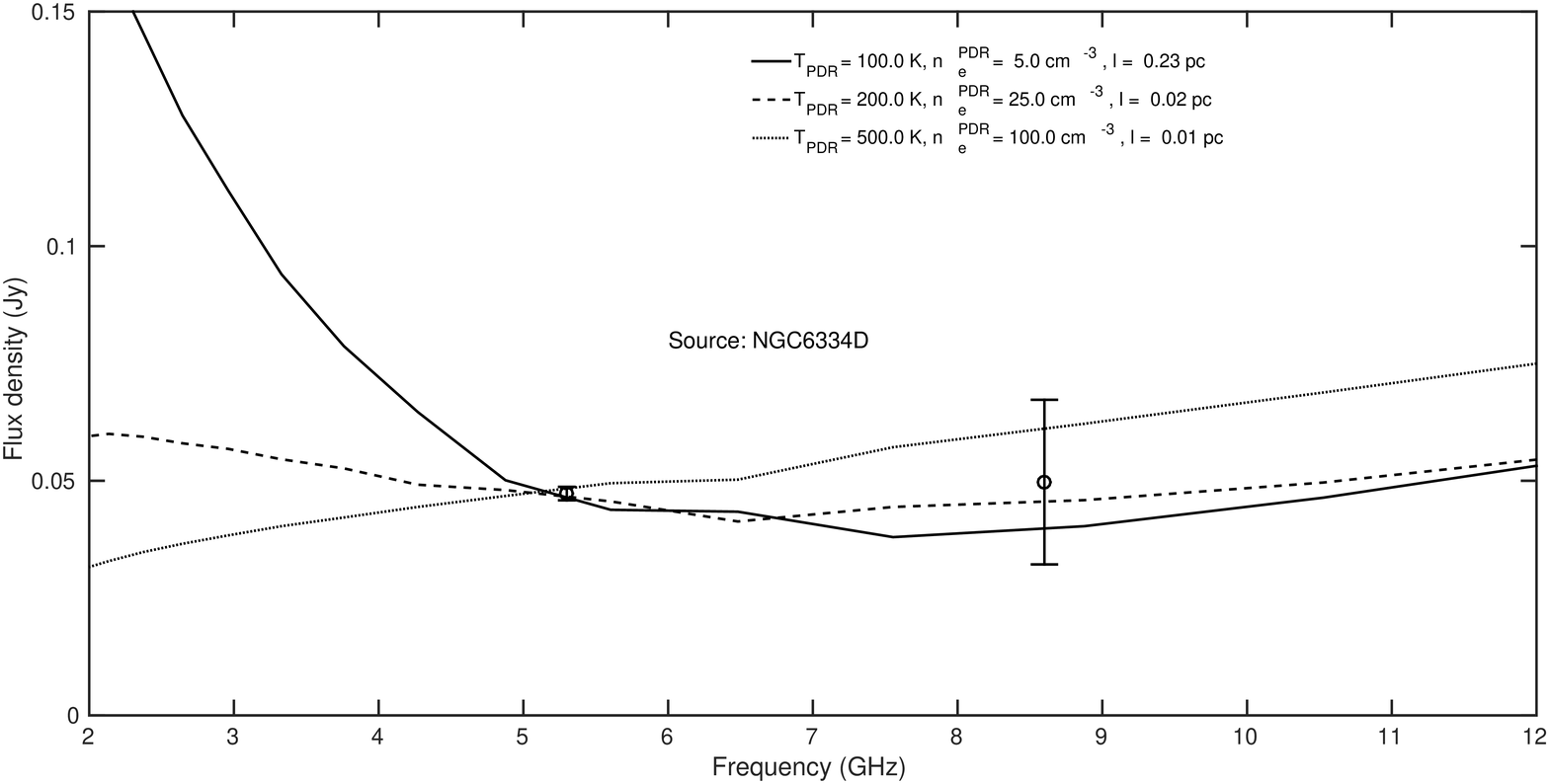} 
\caption{PDR model results for \ngc{6334D}.  See Figure~\ref{fig:model_w3} for
  details.}
\label{fig:model_n6334d}
\end{figure}

For each PDR temperature we ran a grid of models with a set of PDR
electron densities, and then solved for the PDR cylinder thickness.
So for each ($T_{\rm PDR}$, $n_{\rm e}^{\rm PDR}$) pair choice, $\ell$
was varied to determine, by eye, the range of $\ell$ that was
consistent to with our two observational data points within the
errors.  Therefore for each PDR temperature we determined a range of
possible values for $n_{\rm e}^{\rm PDR}$ and $\ell$. We explored
$n_{\rm e}^{\rm PDR}$ = 1, 5, 10, 25, and 50\percc\ for W3 and
\ngc{6334A}; and $n_{\rm e}^{\rm PDR}$ = 5, 10, 25, 50, 100, and
200\percc\ for W49 and \ngc{6334D}.  Model results are shown in
Figures~\ref{fig:model_w3} through \ref{fig:model_n6334d} where we
plot the flux density as a function of frequency.  The curves
correspond to the models that set the extreme range in $n_{\rm e}^{\rm
  PDR}$ and $\ell$ for each PDR temperature, whereas the points are
the constraints from the GBT and 140 Foot observations.  For
\ngc{6334A}, only one model in our grid is consistent with the data to
within the uncertainties.  Our modeling predicts lower PDR
temperatures ($\lsim 200$) for W3 and \ngc{6334A}.  The flux density
uncertainties are significantly higher for the 140 Foot X-band data
and therefore dominate the scatter in these plots.  The results are
summarized in Table~\ref{tab:result}.  For each PDR temperature we
show the range of $n_{\rm e}^{\rm PDR}$ and $\ell$ values that ``fit''
the data; that is, their model curves lie within the observed error
bars.  Listed in Table~\ref{tab:result} are the source name, the PDR
temperature, the range of PDR electron densities, the range of
cylinder thicknesses, and the range of magnetic field strengths (see
below).

\section{Magnetic Field Strength}\label{sec:b}

It is now well established that the observed spectral line widths from
molecular clouds are significantly larger than expected from thermal
broadening alone.  \citet{arons75} first proposed that this
non-thermal line width is due to MHD waves, but the contribution of a
pure hydrodynamic turbulence component cannot be ruled out
\citep{morris74}.  Spectral line widths from PDRs, which reside at the
interface between the molecular cloud and the \hii\ region, are also
dominated by non-thermal broadening. \citet{roshi07} investigated the
origin of this non-thermal component in PDRs using carbon RRLs.  He
concluded that (1) the origin of the non-thermal carbon RRL width is
magnetic; (2) the non-thermal line width is approximately the
Alf$\acute{v}$en speed in the PDR; and (3) the minimum MHD wavelength
for which carbon ions and neutrals are strongly coupled is much
smaller than the size of the PDR.

Perturbations in the magnetic field due to MHD waves create a
velocity field in the plasma. This velocity field results in the
non-thermal broadening of the observed spectral lines. The amplitude
of the velocity field will be equal to the Alf$\acute{v}$en speed if
the perturbing magnetic field is approximately equal to the total
magnetic field strength, B. But pure hydrodynamic motions in the PDR
may also contribute to the non-thermal width of spectra lines.  We
therefore introduce a parameter $\alpha$ to relate the
Alf$\acute{v}$en speed, $V_{\rm A}$ and the observed non-thermal width
of the spectral line:
\begin{equation}\label{eq:va}
V_{\rm A} = \alpha \frac{\Delta{V_{\rm nt}}}{\sqrt{8 ln(2)}},
\end{equation}
where $\Delta{V_{\rm nt}}$ is the FWHM non-thermal line width
defined as
\begin{equation}\label{eq:nonthermal}   
\Delta{V_{\rm nt}} = \sqrt{\Delta{V}^2 - \Delta{V_{\rm t}}^2}.
\end{equation}
Here $\Delta{V}$ is the observed FWHM line width and
$\Delta{V_{\rm t}}$ is the thermally broadened FWHM line width
given by
\begin{equation}\label{eq:thermal}
\Delta{V_{\rm t}} = \bigg[4\,ln(2)\bigg(\frac{2kT_{\rm PDR}}{m_{\rm c}}\bigg) \bigg]^{1/2},
\end{equation}
where $m_{\rm c}$ is the mass of the carbon atom \citep[][Equation
57]{shaver75}.  At one extreme, $\alpha \sim 0$ if the turbulence is
non-magnetic in origin.  At the other extreme, where the magnetic
field dominates the turbulent motions, $\alpha \le \sqrt{3}$.  The
exact value depends on the geometry of the magnetic and matter
perturbations in the PDR since we need to convert the observed, one
dimensional velocity dispersion into a three dimensional velocity
dispersion \citep[see for example][]{mckee95}.  The parameter $\alpha$
must be determined by observations.  \citet{roshi07} compared the
magnetic field strength measured via the Zeeman effect in molecular
clouds with the magnetic field strength derived from carbon RRLs in
PDRs (see Roshi's Figure~3).  Such a comparison is possible since it
has been shown that the magnetic field strength scales with density
\citep{crutcher99}.  From this comparison \citet{roshi07} concluded
that the non-thermal motions in PDRs are primarily caused by MHD waves
and that $\alpha \sim 1$.  Here we follow \citet{roshi14} and take
$\alpha = \sqrt{3}/2 = 0.87$ as a mean value between the two extremes,
mentioned above.

The magnetic field is given by
\begin{equation}\label{eq:b}   
B = V_{\rm A} \sqrt{4\pi\,\rho},
\end{equation}
where $\rho$ is the mass density of the gas coupled to the magnetic
field \citep{parker79}.  \citet{roshi07} showed that the size of the
PDR is larger than the minimum MHD wavelength.  Thus in the PDR there
exists a spectrum of MHD waves for which carbon ions and neutrals are
strongly coupled to the field. The perturbations produced by these
waves result in the non-thermal broadening of carbon lines.  The ions
and neutrals are coupled to these waves which implies that $\rho$ in
Equation~\ref{eq:b} should be the total (i.e., ion, atomic, and
molecular) mass density of the PDR.  The PDR mass density is given by
\begin{equation}
\rho = n_{\rm H} \, \mu \, m_{\rm H},
\end{equation}
where $n_{\rm H}$ is the hydrogen number density, $m_{\rm H}$ is the
hydrogen mass, and $\mu$ is the mean molecular weight.  The hydrogen
number density is determined by modeling the observed carbon lines.  A
pure hydrogen and helium gas with a He/H ratio of 10\% by number
yields $\mu = 1.4$.  Since the contribution of heavier elements, such
as carbon, to the mean molecular weight is negligible, we take $\mu =
1.4$.

In Table~\ref{tab:result} we list a range of magnetic field strengths
calculated using Equation~\ref{eq:b} and the range of determined
$n_{\rm e}^{\rm PDR}$ values for each PDR temperature.  Most of the
uncertainty in determining $B$ comes from our PDR models and therefore
we specify a range of possible values instead of a value and
$1\,\sigma$ error.  The exception is \ngc{6334A} where only one model
fits the data.  The magnetic field strength for W49 and \ngc{6334D}
are not well constrained with a wide range of possible magnetic field
strengths.  N.B., these uncertainties do not include any contribution
to $\alpha$ which can range from 0 to $\sqrt{3}$.

\section{Discussion}\label{sec:disc}

The role of magnetic fields in star formation has been an important
astrophysical topic for decades \citep{shu87, mckee07}.  Recently, the
debate has centered around competing theories that depend on the
strength of the magnetic field.  Strong magnetic fields will support
molecular clouds from collapse, but neutral material will slip past
these fields and thereby increase the molecular mass in a process
called ambipolar diffusion \citep{shu87}.  On the other hand, weak
magnetic fields allow molecular clouds to form from turbulent flows on
time scales equal to the free fall time \citep{maclow04}.  There are
also star formation theories that include both of these processes
\citep[e.g.,][]{nakamura05}.  Understanding magnetic field properties
in star forming complexes provides important constraints to these
theories \citep{crutcher12}.

Measuring magnetic field properties in star forming complexes is
difficult, however, and therefore additional data are necessary to
properly constrain star formation models.  \citet{roshi07} proposed a
new method of deriving the magnetic field strength in PDRs using
carbon RRLs.  If the non-thermal motions in PDRs are dominated by MHD
waves, then the non-thermal line widths provide a measure of the
magnetic field strength.  Here we test this hypothesis by comparing
the magnetic field strength derived from carbon RRLs with Zeeman
measurements in four sources: W3, W49, \ngc{6334A} and \ngc{6334D}.
Zeeman observations provide a measure of the magnetic field along the
line-of-sight (LOS) and therefore information on the morphology and a
lower limit to the magnetic field strength (i.e., $|B_{\rm los}| \le\
B$).  If we consider a large number of PDRs that have a magnetic field
that is oriented randomly relative to the LOS, then statistically $B =
2\,|B_{\rm los}|$ \citep{crutcher99}.  The magnetic field orientations
in PDRs may not be random, however, given their geometry and
formation.

Below we discuss each source separately.

W3 is a nearby, bright \hii\ region with at least 8 resolved
components (A-H) in the core region \citep[see, e.g.,][]{tieftrunk97}.
\citet{roberts93} derived the LOS magnetic field strength from \hi\
Zeeman observations to be $B_{\rm los} = -47 \pm\ 3$\microgauss, $+103
\pm\ 7$\microgauss, and $+36 \pm\ 6$\microgauss\ toward components A,
B, and C+D, respectively.  Our GBT observations are centered near W3A
but include most of the core.  We therefore consider $|B_{\rm los}| =
30 - 110$\microgauss\ for comparison with our GBT data.  The LSR
velocity of the carbon RRL is $-39.1$\kms, consistent with the \hi\
Zeeman data.  We estimate $B$ to be between $140-320$\microgauss.

W49 is one of the most luminous star forming complexes in the Galaxy
and contains the \hii\ region W49A and a supernova remnant W49B
\citep{depree97}.  Our GBT observations cover the W49A north region
that consists of a ring of ultracompact \hii\
regions. \citet{brogan01} detected the \hi\ Zeeman line toward W49A
and determined $|B_{\rm los}| = 60-300$\microgauss.  The $V_{\rm LSR}
\sim 4$\kms\ $B_{\rm los}$ component is negative, whereas the $V_{\rm
  LSR} \sim 7$\kms\ $B_{\rm los}$ component is positive.  Also, higher
resolution Zeeman detections are stronger.  Our GBT data constrain the
magnetic field strength to be between $B = 200-1300$\microgauss.  Our
C-band data contain two velocity components ($\sim 14$\kms\ and $\sim
4$\kms), but the lower spectral resolution 140 Foot, X-band data has
only one component at $\sim 7$\kms.  We have modeled the PDR using the
C-band $\sim 14$\kms\ component.  The results are similar if we use
the $\sim 4$\kms\ component.

Since W49 is complex and distant ($R_{\rm Sun} = 11.8$\kpc), it is
difficult to compare the VLA Zeeman observations with our lower
resolution GBT data.  The carbon RRL emission regions may not be
probing the same volume of gas as the VLA \hi\ data.  The differences
in LSR velocity between our C-band and X-band carbon RRL data are
troubling and therefore our results for W49 are suspect.  Furthermore,
the carbon RRL intensity is weighted by the emission measure
($\propto n_{\rm e}^{2}$), whereas the \hi\ intensity is proportional
to the hydrogen column density.  So the carbon RRLs may be probing
denser gas where the magnetic field strengths should be higher.

\ngc{6334} is a nearby ($R_{\rm Sun} = 1.7$\kpc), star forming region
that contains at least seven star forming components
\citep{kraemer00}.  Here we focus on components A and D.
\citet{sarma00} made \hi\ and OH VLA measurements toward both of these
components with LSR velocities around $-2$\kms\ to $-5$\kms,
consistent with our carbon RRL velocities. Significant Zeeman
detections were made toward \ngc{6334A} in OH where $B_{\rm los} = 148
\pm\ 20$\microgauss\ and $B_{\rm los} = 162 \pm\ 33$\microgauss\ for
the 1665\mhz\ and 1667\mhz\ lines.  We consider $|B_{\rm los}| =
128-195$\microgauss\ for comparison with GBT data.  From our carbon
RRLs we estimate $B = 190 \pm\ 96$\microgauss, where we assume a 50\%
error.  For \ngc{6334D}, \citet{sarma00} measure $B_{\rm los} = -93
\pm\ 13$\microgauss\ from \hi\ Zeeman spectra yielding $|B_{\rm los}|
= 80-106$.  Our carbon RRL data toward \ngc{6334D} do not provide a
good constraint for the magnetic field with $B = 180-1200$\microgauss,
but the lower bound is within a factor of two relative to the Zeeman
value.

\begin{figure}
\includegraphics[angle=0,scale=0.8]{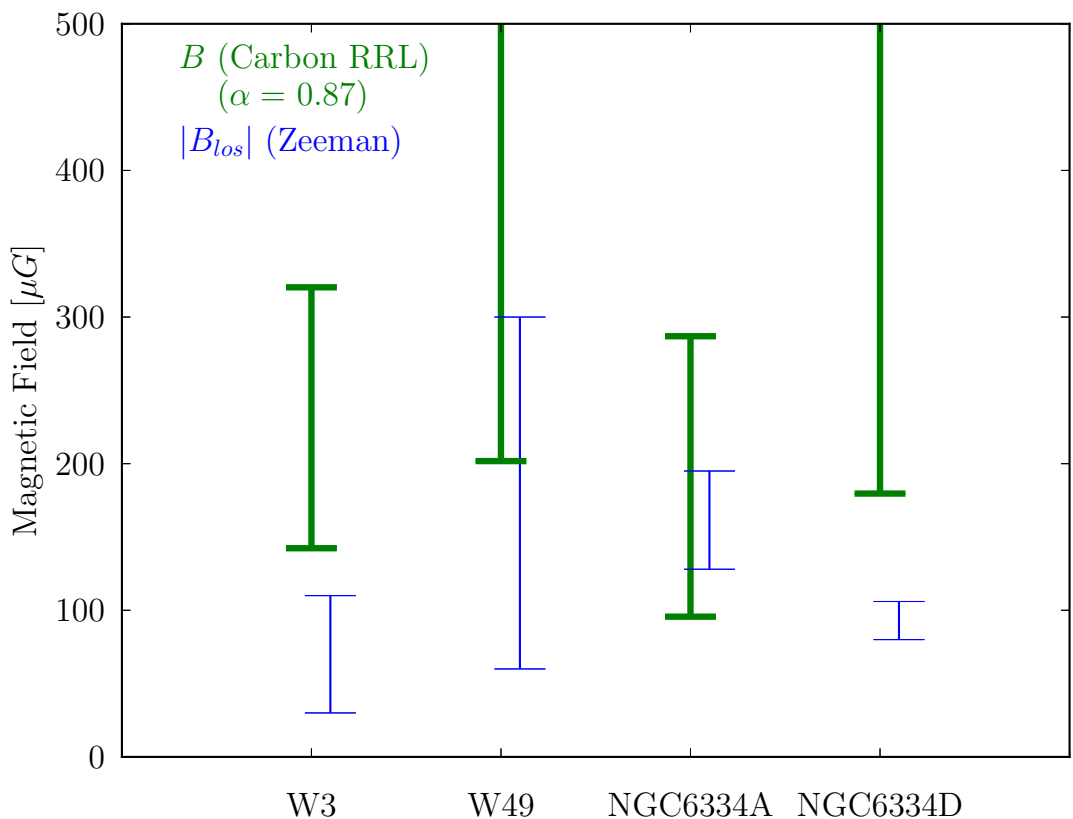} 
\caption{Magnetic field strengths derived here from carbon RRLs (thick
  green lines) for W3, W49, \ngc{6334A}, and \ngc{6334D}, compared to
  line-of-sight magnetic field strengths from Zeeman observations
  (blue lines).  The magnetic field strengths from carbon RRLs
  correspond to the range of values in our model grid that are
  consistent with the data.  The range for $|B_{los}|$ are taken from
  \hi\ or OH Zeeman measurements in the literature (see text).  The
  plot has been truncated at $B = 500\,\mu G$ for clarity.}
\label{fig:b}
\end{figure}

Figure~\ref{fig:b} summarizes these results.  We plot a comparison of
the magnetic field strength from our carbon RRL measurements with the
LOS magnetic field strength from Zeeman spectra as discussed above.
We assume a 50\% error for \ngc{6334A} since we were not able to
determine a range of $B$ values from our models.  The uncertainties
are quite large but the trend is that $|B_{\rm los}|$ is between
$0.5-1.0$ of the lower bound of $B$, and this is consistent with our
expectations that $|B_{\rm los}| \le\ B$.  Similar results are
obtained for PDRs in OrionB \citep{roshi14} and W3, W49, and S88B
\citep{roshi07}.

Overall, our GBT carbon RRL data are consistent with the hypothesis of
\citet{roshi07} that the non-thermal motions in PDRs have a magnetic
origin.  But our results are not conclusive since we do not know
$\alpha$ or the orientation of the magnetic field vector.  Moreover,
there are many assumptions and approximations in deriving the magnetic
field strength using the carbon RRL method.  Here we list some of the
issues.

\begin{enumerate}

\item {\it PDR Geometry:} We assume the PDR region is a thin cylinder
  that covers the \hii\ region.  Infrared data do show that PDR
  material as seen in 8\microns\ emission typically surrounds \hii\
  regions (24\microns\ emission) with a thin, sheet-like morphology
  \citep[e.g.,][]{anderson14}.  Our GBT and 140 Foot data lack the
  spatial resolution, however, to confirm that the PDR covers the
  \hii\ region.  Interferometers like the VLA can spatially resolve
  some of these regions with enough sensitivity to verify this
  geometry \citep[see][]{roshi14}.

\item {\it Model Constraints:} Observations at only two frequencies
  are used to constrain three free parameters: $T_{\rm PDR}$, $n_{\rm
    e}^{\rm PDR}$, and $\ell$.  Therefore we had to assume several
  values for the PDR temperature to constrain the fits.  This could be
  significantly improved by obtaining additional carbon RRL data
  separated in frequency.  

\item {\it \hi/OH Zeeman Data:} We use \hi\ and OH Zeeman data to
  check the hypothesis by \citet{roshi07} that carbon RRL non-thermal
  widths are magnetic in origin with the goal of using such data to
  determine magnetic field strengths in PDRs.  We expect the \hi\ and
  OH emission to arise from the PDR but this emission may not be
  sampling the same region as the carbon RRL emission.  RRLs probe
  higher density gas compared with \hi, and our models indicate that
  cm-wavelength carbon RRLs are sensitive to PDR material in front of
  the \hii\ region relative to our line-of-sight.  So we may not be
  sampling the same material.  This can be mitigated by observing the
  carbon RRL Zeeman effect.

\item {\it Alf$\acute{\it v}$en Speed:} We calculate $V_{\rm A}$ from
  $\Delta{V_{\rm nt}}$.  If the gas pressure is small compared to the
  magnetic pressure then the velocity dispersion should be
  approximately the Alf$\acute{v}$en speed \citep[see][]{arons75}.
  But we have to convert the one-dimensional velocity dispersion,
  $\Delta{V_{\rm nt}}$, to a three-dimensional velocity dispersion.
  Since we do not know the magnetic field geometry the value of
  $\alpha$ must be constrained from observations.  Therefore, the
  parameter $\alpha$ in Equation~\ref{eq:va} is another free
  parameter.  The magnetic field strength is proportional to $\alpha$,
  and here we assume $\alpha = \sqrt{3}/2$.  The uncertainties given
  in Table~\ref{tab:result} are taken from the PDR models and do not
  include the uncertainties in $\alpha$.

\item {\it B versus B$_{\it los}$:} Since Zeeman observations probe
  the LOS magnetic field strength we cannot directly compare these
  results with our estimates of the total magnetic field strength from
  our carbon RRL data for a given source.  If we observed many PDRs
  using both methods we could make a statistical argument that $B =
  2\,|B_{\rm los}|$ \citep[e.g.,][]{crutcher99}.  But this assumes
  that the orientation of the magnetic field vector is random which
  may not be true for PDRs.

\end{enumerate}

How to proceed?  Observations of carbon RRLs at several different
frequencies using both the VLA and GBT could significantly improve our
understanding of the PDR geometry and provide better constraints to
the models.  Observing many sources would allow a statistical
comparison with Zeeman results and an estimate of $\alpha$.  A more
direct comparison of the magnetic field strength could be made by
measuring the Zeeman effect in carbon RRLs.  To do this in many
sources with good accuracy, however, would probably require the SKA or
NGVLA. Nevertheless, our results here are consistent with the
\citet{roshi07} hypothesis of a magnetic origin for the observed
carbon RRL non-thermal line widths.

\section{Summary}

Magnetic fields play an important role in star formation, but they are
difficult to measure, and therefore have not provided very stringent
constraints on a host of relevant astrophysical processes.
\citet{roshi07} proposed a new technique to derive magnetic field
strengths using carbon RRLs in PDRs.  It assumes that the non-thermal
motions in PDRs are dominated by MHD waves.  Here we measure the
\cc104-\cc110\ (5.3\ghz) RRL emission with the GBT toward four star
forming regions to test this hypothesis.  We use the models developed
by \citet{roshi14} to calculate the carbon RRL flux density by
performing the non-LTE radiative transfer.  To constrain these models
requires at least two carbon RRLs separated in frequency, and to do
this we use the \cx91-\cx92\ (8.7\ghz) RRLs from \citet{quireza06}
together with the observations reported here.

We estimate $B \sim 100-300$\microgauss\ in W3 and \ngc{6334A}, and $B
\sim 200-1000$\microgauss\ in W49 and \ngc{6334D}.  These results are
consistent with \hi\ and OH Zeeman observations, which measure the
line-of-sight magnetic field strength $B_{\rm los}$.  That we find
$|B_{\rm los}| \le\ B$ in all cases is consistent with the hypothesis
that the non-thermal component of the velocity dispersion measured by
carbon RRLs in magnetic in origin.  There are many assumptions and
approximations made in deriving $B$, however, and therefore to use
this method to determine magnetic field strengths accurately may
require telescopes like the SKA or NGVLA.

\acknowledgements

{\it Facility:} \facility{GBT}

\clearpage

%
%

%
%

\clearpage

\begin{deluxetable}{lccrcccccccc}
\tabletypesize{\scriptsize}
\tablecaption{\hii\ Region Continuum Parameters \label{tab:cont}}
\tablewidth{0pt}
\tablehead{
\colhead{} & \colhead{} & \colhead{} & \colhead{} &
\multicolumn{4}{c}{\underline{~~~~~~~~~~~~~~~~~~~~R.A.~~~~~~~~~~~~~~~~~~~~}} & 
\multicolumn{4}{c}{\underline{~~~~~~~~~~~~~~~~~~~~Decl.~~~~~~~~~~~~~~~~~~~~}} \\ 
\colhead{} & \colhead{R.A. (B1950)} & \colhead{Decl. (B1950)} &
\colhead{$R_{\rm Sun}$} & 
\colhead{$T_{\rm C}$} & \colhead{$\sigma\,T_{\rm C}$} & 
\colhead{$\Theta$} & \colhead{$\sigma\,\Theta$} & 
\colhead{$T_{\rm C}$} & \colhead{$\sigma\,T_{\rm C}$} & 
\colhead{$\Theta$} & \colhead{$\sigma\,\Theta$} \\ 
\colhead{Name} & \colhead{(hh:mm:ss.s)} & \colhead{(dd:mm:ss)} &
\colhead{(kpc)} & 
\colhead{(K)} & \colhead{(K)} & \colhead{(arcsec)} & \colhead{(arcsec)} & 
\colhead{(K)} & \colhead{(K)} & \colhead{(arcsec)} & \colhead{(arcsec)}
} 
\startdata 
W3 A      & 02:21:56.9 &  $+$61:52:40.0 &  2.1\tablenotemark{a} & 70.42 & 0.35 & 360.25 & 2.14 & 69.34 & 0.19 & 154.37 & 0.62 \\
W49 A     & 19:07:52.1 &  $+$09:01:08.0 & 11.8\tablenotemark{a} & 58.66 & 0.57 & 202.66 & 2.38 & 59.53 & 0.15 & 161.33 & 0.53 \\
NGC6334 A & 17:16:57.8 &  $-$35:51:45.0 &  1.7\tablenotemark{b} & 34.37 & 0.11 & 208.69 & 1.06 & 34.48 & 0.17 & 182.41 & 1.84 \\
NGC6334 D & 17:17:23.0 &  $-$35:46:20.0 &  1.7\tablenotemark{b} & 39.91 & 0.16 & 246.70 & 1.86 & 38.20 & 0.10 & 189.84 & 0.81 \\
\enddata 
\tablenotetext{a}{From \citet{bania97}.}
\tablenotetext{b}{From \citet{neckel78}.}
\end{deluxetable}

\begin{deluxetable}{lcrrrrrrrr}
\tabletypesize{\scriptsize}
\tablecaption{Radio Recombination Line Parameters \label{tab:line}}
\tablewidth{0pt}
\tablehead{
\colhead{} & \colhead{} & \colhead{$T_{\rm L}$} & \colhead{$\sigma\,T_{\rm L}$} & 
\colhead{$\Delta{V}$} & \colhead{$\sigma\,\Delta{V}$} & 
\colhead{$V_{\rm LSR}$} & \colhead{$\sigma\,V_{\rm LSR}$} & 
\colhead{$t_{\rm intg}$} & \colhead{rms\tablenotemark{b}} \\ 
\colhead{Name} & \colhead{Element\tablenotemark{a}} & \colhead{(mK)} & \colhead{(mK)} & 
\colhead{(\kms)} & \colhead{(\kms)} & \colhead{(\kms)} & \colhead{(\kms)} & 
\colhead{(hr)} & \colhead{(mK)}
} 
\startdata 

W3         & H  & 3159.30 & 6.38 & 29.795 & 0.040 & $-$40.462 & 0.010 & 10.7 & 11.20 \\
\nodata    & H  &  416.86 & 6.41 &  9.076 & 0.139 & $-$41.069 & 0.037 & \nodata & 11.20 \\
\nodata    & He &  326.76 & 1.23 & 24.848 & 0.113 & $-$40.222 & 0.046 & \nodata & 8.42 \\
\nodata    & C  &  342.99 & 2.73 &  5.525 & 0.077 & $-$39.054 & 0.035 & \nodata & 8.42 \\
\nodata    & X\tablenotemark{c}  &   80.08 & 2.45 &  6.519 & 0.400 & $-$46.556 & 0.163 & \nodata & 8.42 \\
 & & & & & & & & & \\
W49        & H  & 2918.63 & 2.77 & 29.159 & 0.032 &     8.345 & 0.014 & 29.6 & 12.23 \\
\nodata    & He &  296.59 & 0.74 & 24.039 & 0.084 &     8.857 & 0.032 & \nodata & 4.77 \\
\nodata    & C  &   93.06 & 3.29 &  7.856 & 0.307 &    13.580 & 0.192 & \nodata & 4.77 \\
\nodata    & C  &   96.66 & 1.47 &  9.608 & 0.536 &     4.464 & 0.203 & \nodata & 4.77 \\ 
\nodata    & X\tablenotemark{c}  &   11.35 & 1.59 &  5.805 & 1.43  &  $-$6.386 & 0.539 & \nodata & 4.77 \\
 & & & & & & & & & \\
NGC6334 A  & H  & 2130.88 & 1.54 & 26.827 & 0.023 &     1.035 & 0.009 & 21.4 & 6.75 \\
\nodata    & He &  125.76 & 0.54 & 23.834 & 0.118 &     2.671 & 0.050 & \nodata & 4.44 \\
\nodata    & C  &  184.97 & 1.17 &  5.212 & 0.042 &  $-$2.478 & 0.017 & \nodata & 4.44 \\
\nodata    & X\tablenotemark{d}  &   39.97 & 1.20 &  5.083 & 0.204 & $-$11.141 & 0.077 & \nodata & 4.44 \\
 & & & & & & & & & \\
NGC6334 D  & H  & 3083.08 & 1.63 & 22.270 & 0.014 &  $-$3.375 & 0.006 & 23.7 & 7.59 \\
\nodata    & He &  280.85 & 0.84 & 15.999 & 0.062 &  $-$3.713 & 0.024 & \nodata & 4.73 \\
\nodata    & C  &   94.54 & 0.96 &  7.022 & 0.112 &  $-$2.722 & 0.040 & \nodata & 4.86 \\
\nodata    & X\tablenotemark{d}  &   20.03 & 1.08 &  5.492 & 0.509 & $-$11.888 & 0.172 & \nodata & 4.86 \\

\enddata 
\tablecomments{Spectral line parameters correspond to the average of 7 RRLs
  ($104\alpha-110\alpha$).}
\tablenotetext{a}{The RRL frequencies are specified using the Rydberg
  equation which depends on the reduced mass \citep{gordon09}.}
\tablenotetext{b}{The H RRL was fit separately from the He and heavy
  element RRLs and therefore has a different rms line-free spectral
  noise.}  
\tablenotetext{c}{The line identification is unclear. It may be carbon from
  a different PDR component or possibly sulfur.}
\tablenotetext{d}{The line appears to be sulfur based on the LSR
  velocity and reduced line intensity relative to carbon.}

\end{deluxetable}

\begin{deluxetable}{lccrccc}
\tablecaption{Spherical Homogeneous \hii\ Region Models\tablenotemark{a} \label{tab:hii}}
\tablewidth{0pt}
\tablehead{
\colhead{} & \colhead{$T_{\rm e}$\tablenotemark{b}} & \colhead{$\theta_{\rm sph}$} & 
\colhead{$n_{\rm e}$} &  \colhead{Peak $EM$\tablenotemark{c}} & 
\colhead{Log$_{10}$($N_{\rm L}$)} & \colhead{Spectral\tablenotemark{d}} \\ 
\colhead{Name} & \colhead{($K$)} & \colhead{(arcmin)} & 
\colhead{(cm$^{-3}$)} & \colhead{$10^{5}\,$(pc$\,$cm$^{-6}$)} & 
\colhead{(photons$\,$s$^{-1}$)} & \colhead{Type}
}
\startdata 

W3 A         & 8000 & 5.24 &  522 & 7.23 & 49.63 & O4.5 \\
W49 A        & 8500 & 3.26 &  318 & 6.22 & 50.82 & $<$O3 \\
\ngc{6334} A & 8000 & 3.89 &  526 & 3.54 & 48.97 & O7.5 \\
\ngc{6334} D & 7000 & 4.56 &  480 & 3.84 & 49.13 & O7 \\

\enddata 
\tablenotetext{a}{See \citet{balser95}.}
\tablenotetext{b}{Taken from \citet{balser99}.}
\tablenotetext{c}{See \citet{wood89}.}
\tablenotetext{d}{Using the stellar models of \citet{vacca96}.}
\end{deluxetable}

\begin{deluxetable}{lcrrrrrrrr}
\tabletypesize{\scriptsize}
\tablecaption{Radio Recombination Line PDR Model Constraints \label{tab:model}}
\tablewidth{0pt}
\tablehead{
\colhead{} & \colhead{} & \colhead{$T_{\rm L}$} & \colhead{$\sigma\,T_{\rm L}$} & 
\colhead{$\Delta{V}$} & \colhead{$\sigma\,\Delta{V}$} & 
\colhead{$V_{\rm LSR}$} & \colhead{$\sigma\,V_{\rm LSR}$} & 
\colhead{$S_{\nu}$} & \colhead{$\sigma\,S_{\nu}$} \\ 
\colhead{Name} & \colhead{RRL\tablenotemark{a}} & \colhead{(mK)} & \colhead{(mK)} & 
\colhead{(\kms)} & \colhead{(\kms)} & \colhead{(\kms)} & \colhead{(\kms)} & 
\colhead{(mJy)} & \colhead{(mJy)}
}
\startdata 
W3 A         & $<$\cc104-\cc110$>$ & 342.99 & 2.73  &  5.525 & 0.077  &  $-$39.054 & 0.035 & 171.5 & 1.4 \\
             & \cx91\              &  56.39 & 4.93  &  7.68  & 0.79   &  $-$40.10  & 0.33  & 113.4 & 9.9 \\
\tableline
W49 A        & $<$\cc104-\cc110$>$ &  93.06 & 3.29  &  7.856 & 0.307  &     13.58  & 0.192  &  46.5 & 1.6 \\
             & $<$\cc104-\cc110$>$ &  96.66 & 1.47  &  9.608 & 0.536  &      4.464 & 0.203  &  48.3 & 0.7 \\
             & $<$\cx91-\cx92$>$   &  24.77 & 1.79  & 15.72  & 1.36   &      7.33  & 0.67   &  49.8 & 3.6 \\
\tableline
\ngc{6334} A & $<$\cc104-\cc110$>$ & 184.97 &  1.17 &  5.221 & 0.042  &   $-$2.478 & 0.017  &  92.5 & 0.6 \\
             & \cx91\              &  26.93 &  3.42 &  8.34  & 1.23   &   $-$3.14  & 0.52   &  54.1 & 6.9 \\
\tableline
\ngc{6334} D & $<$\cc104-\cc110$>$ &  94.54  & 0.96 &  7.022 & 0.112  &   $-$2.722 & 0.040  &  47.3 & 0.5 \\
             & \cx91\              &  24.73  & 2.91 &  9.99  & 1.37   &   $-$4.37  & 0.58   &  49.7 & 5.8  \\

\enddata 


\tablenotetext{a}{The C-band (\cc104-\cc110) RRL data are from the GBT
  (Table~\ref{tab:line}), and the X-band (\cx91-\cx92) RRL data are
  from the 140 Foot telescope \citep{quireza06}.}



\end{deluxetable}

\begin{deluxetable}{lcccc}
\tablecaption{PDR Model Physical Properties \label{tab:result}}
\tablewidth{0pt}
\tablehead{
\colhead{} & \colhead{$T_{\rm PDR}$} & 
\colhead{$n_{\rm e}^{\rm PDR}$} &
\colhead{$\ell$} & \colhead{$B$\tablenotemark{a}}  \\
\colhead{Name} & \colhead{(K)} & \colhead{(\percc)} & 
\colhead{(pc)}& \colhead{($\mu$G)}
}
\startdata 
W3 A     & 100  & $5-25$  & $0.4-0.013$ & $140-320$  \\
         & 200  & $5-25$  & $1.7-0.048$ & $140-320$ \\
\tableline
W49 A    & 100  & $50-100$  & $0.0015-0.0004$ & $650-910$  \\
         & 200  & $25-200$  & $0.0199-0.0004$ & $460-1300$ \\
         & 500  & $5-200$   & $3.1-0.0016$    & $200-1300$ \\
\tableline
\ngc{6334} A & 100  & 10  & 0.07 & 190  \\
\tableline
\ngc{6334} D & 100  & $5-100$  & $0.23-0.0005$ & $180-820$ \\
             & 200  & $5-200$  & $0.73-0.0004$ & $180-1200$ \\
             & 500  & $5-200$  & $3.4-0.0017$  & $180-1100$ \\
\enddata 

\tablenotetext{a}{The range in $B$ is determined from the PDR models
  and does not include any uncertainty in $\alpha$ (see
  \S{\ref{sec:b}}).}
\end{deluxetable}


\begin{thebibliography} {}

\bibitem[Abel et al.(2005)]{abel05} Abel, N. P., Ferland, G. J., Shaw,
  G., \& van Hoof, P. A. M. 2005, \apjs, 161, 65

\bibitem[Altenhoff(1960)]{altenhoff60} Altenhoff, W. J. 1960,
  Ver\"{o}ff Sternwarte Bonn, Nr. 59

\bibitem[Anderson et al.(2014)]{anderson14} Anderson, L. D., Bania, T.
  M., Balser, D. S., et al. 2014, \apjs, 212, 1

\bibitem[Arons \& Max(1975)]{arons75} Arons, J., \& Max, C. E. 1975,
  \apj, 196, L77

\bibitem[Balser(2006)]{balser06} Balser, D. S. 2006, \aj, 132, 2326

\bibitem[Balser et al.(1995)]{balser95} Balser, D. S., Bania, T. M.,
  Rood, R. T., \& Wilson, T. L. 1995, \apjs, 100, 371

\bibitem[Balser et al.(1999)]{balser99} Balser, D. S., Bania, T. M.,
  Rood, R. T., \& Wilson, T. L. 1999, \apj, 510, 759

\bibitem[Bania et al.(1997)]{bania97} Bania, T. M., Balser, D. S.,
  Rood, R. T., Wilson, T. L., \& Wilson, T. J.  1997, \apjs, 113, 353

\bibitem[Beckman \& Rela$\tilde{\rm n}$o(2004)]{beckman04} Beckman,
  J. E., \& Rela$\tilde{\rm n}$o, M. 2004, Ap\&SS, 292, 111

\bibitem[Brocklehurst \& Salem(1977)]{brocklehurst77} Brocklehurst,
  M., \& Salem, M. 1977, CoPhC, 13, 39

\bibitem[Brogan \& Troland(2001)]{brogan01} Brogan, C. L., \& Troland,
  T. H. 2001, \apj, 560, 821

\bibitem[Crutcher(1999)]{crutcher99} Crutcher, R. M. 1999, \apj, 520,
  706.

\bibitem[Crutcher(2012)]{crutcher12} Crutcher, R. M. 2012, \araa, 50,
  29.

\bibitem[De Pree et al.(1997)]{depree97} De Pree, C. G., Mehringer,
  D. M., \& Goss, W. M. 1997, \apj, 482, 307

\bibitem[Ferland(2001)]{ferland01}
Ferland, G. 2001, \pasp, 113, 41

\bibitem[Gordon \& Sorochenko(2009)]{gordon09} Gordon, M. A., \&
  Sorochenko, R. L. 2009, Radio Recombination Lines. Their Physics and
  Astronomical Applications (ASSL, Vol. 282; New York: Springer)

\bibitem[Kraemer et al.(2000)]{kraemer00} Kraemer, K., E., Jackson,
  J. M., Lane, A. P., \& Paglione, T. A. D. 2000, \apj, 542, 946

\bibitem[MacLow \& Klessen(2004)]{maclow04}
MacLow M.-M., \& Klessen R.S. 2004., RvMP 76, 125

\bibitem[Marscher, et al.(1993)]{marscher93}
Marscher, A. P., Moore, E. M., \& Bania, T. M. 1993, \apj, 419, L101

\bibitem[McKee \& Ostriker(2007)]{mckee07}
McKee C. F., Ostriker E. C. 2007, \araa, 45, 565

\bibitem[McKee \& Zweibel(1995)]{mckee95}
McKee, C. F., \& Zweibel, E. G. 1995, \apj, 440, 686

\bibitem[Minter \& Spangler(1996)]{minter96} Minter, A. H., \&
  Spangler, S. R. 1996, \apj, 458, 194

\bibitem[Morris et al.(1974)]{morris74} Morris, M., Zuckerman, B.,
  Turner, B. E., \& Palmer, P. 1974, \apj, 192, L27

\bibitem[Morton(1974)]{morton74} Morton, D. C. 1974, \apjl, 193, L35

\bibitem[Mouschovias(1975)]{mouschovias75} Mouschovias, T. Ch. 1975,
  Ph.D. thesis, Univ. California, Berkeley

\bibitem[Nakamura \& Li(2005)]{nakamura05} Nakamura F., \& Li
  Z.-Y. 2005, \apj, 631, 411

\bibitem[Natta et al.(1994)]{natta94} Natta, A., Walmsley, C. M., \&
  Tielens, A. G. G. M. 1994, \apj, 428, 209

\bibitem[Neckel(1978)]{neckel78} Neckel, T. 1978, \aap, 69, 51

\bibitem[Parker(1979)]{parker79} Parker, E. N. 1979, Cosmical Magnetic
  Fields (International Series of Monographs on Physics; New York:
  Oxford Univ. Press)

\bibitem[Quireza et al.(2006)]{quireza06} Quireza, C., Rood, R. T.,
  Balser, D. S., \& Bania, T. M. 2006, ApJS, 165, 338

\bibitem[Roberts et al.(1993)]{roberts93} Roberts, D. A., Crutcher,
  R. M., Troland, T. H., \& Goss,W. M. 1993, \apj, 412, 675

\bibitem[Roshi(2007)]{roshi07} Roshi, D. A. 2007, \apjl, 658,
  L41

\bibitem[Roshi et al.(2005)]{roshi05} Roshi, D. A., Balser, D. S.,
  Bania, T. M., Goss, W. M., \& De Pree, C. G. 2005, \apj, 625, 181

\bibitem[Roshi et al.(2006)]{roshi06} Roshi, D. A., De Pree, C. G.,
  Goss, W. M., \& Anantharamaiah, K. R. 2006, \apj, 644, 279

\bibitem[Roshi et al.(2014)]{roshi14}
Roshi, D. A., Goss, W. M., Jeyakumar, S. 2014, \apj, 793, 83

\bibitem[Sarma et al.(2000)]{sarma00} Sarma, A. P., Troland, T. H.,
  Roberts, D. A., \& Crutcher, R. M. 2000, \apj, 533, 271

\bibitem[Shaver(1975)]{shaver75} Shaver, P. A. 1975, Parmana, 5, 1

\bibitem[Shu et al.(1987)]{shu87}
Shu F. H., Adams F. C., Lizano S. 1987, \araa, 25, 23

\bibitem[Tieftrunk et al.(1997)]{tieftrunk97} Tieftrunk, A. R., Gaume,
  R. A., Claussen, M. J.,Wilson, T. L., \& Johnston, K. J.  1997, \aap,
  318, 931

\bibitem[Vacca et al.(1996)]{vacca96} Vacca, W. D., Garmany, C. D., \&
  Shull, J. M. 1996, \apj, 460, 914

\bibitem[Walmsley \& Watson(1982)]{walmsley82} Walmsley, C. M., \&
  Watson, W. D. 1982, \apj, 260, 317

\bibitem[Wenger et al.(2013)]{wenger13} Wenger, T. V., Bania, T. M.,
  Dana, S., \& Anderson, L. D. 2013, \apj, 764, 34

\bibitem[Wood \& Churchwell(1989)]{wood89}
Wood, D. O. S., \& Churchwell, E. 1989, \apjs, 69, 831

\bibitem[Zuckerman \& Palmer(1968)]{zuckerman68}
Zuckerman, B., \& Palmer, P. 1968, \apjl, 153, L145

\end{thebibliography}
\end{document}